\documentclass[
    10pt,
    final,
    journal,
    letterpaper,
    twoside,
    twocolumn,
]{./style/IEEEtran}

\usepackage{epsfig}
\usepackage{subfigure}
\usepackage{amsmath}
\usepackage{amssymb}

\usepackage{psfrag}
\usepackage{cite}
\usepackage{graphicx}
\usepackage{multirow}
\usepackage{array}
\usepackage{enumerate}
\usepackage{graphicx}
\usepackage{times}

\usepackage[ruled,linesnumbered,vlined]{algorithm2e}

\usepackage{balance}   
\usepackage{color}
\usepackage[colorlinks=false, linkcolor=blue, bookmarks=false]{hyperref}

\newcommand{\REV}[1]{\textcolor[rgb]{0.00,0.00,0.00}{{#1}}}

\SetKwFunction{DFS}{DFS}
\SetKwFunction{BFS}{BFS}
\SetKwFunction{sort}{sort}

\begin{document}

\title{On the Entropy Computation of Large Complex Gaussian Mixture Distributions
\author{Su~Min~Kim,~\IEEEmembership{Member,~IEEE}, Tan~Tai~Do,~\IEEEmembership{Student Member,~IEEE}, \\Tobias J. Oechtering,~\IEEEmembership{Senior Member,~IEEE}, and Gunnar Peters}
\thanks{Copyright (c) 2015 IEEE. Personal use of this material is permitted. However, permission to use this material for any other purposes must be obtained from the IEEE by sending a request to pubs-permissions@ieee.org.}
\thanks{Part of this work has been presented at \emph{IEEE Workshop on Statistical Signal Processing (SSP)}, Gold Coast, Australia, 2014.}
\thanks{S. M. Kim was with the School of Electrical Engineering, KTH Royal Institute of Technology, Stockholm, Sweden, and he is currently with the Department of Electronics Engineering, Korea Polytechnic University, Siheung, Korea (e-mail: \texttt{suminkim@kpu.ac.kr}).}
\thanks{T. T. Do and T. J. Oechtering are with the School of Electrical Engineering, KTH Royal Institute of Technology, Stockholm, Sweden (e-mail: \texttt{\{ttdo,oech\}@kth.se}).}
\thanks{G. Peters is with the Huawei Technologies Sweden AB, Stockholm, Sweden (e-mail: \texttt{gunnar.peters@huawei.com}).}
}


\maketitle

\begin{abstract}
The entropy computation of Gaussian mixture distributions with a large number of components has a prohibitive computational complexity.
In this paper, we propose a novel approach exploiting the sphere decoding concept to bound and approximate such entropy terms with reduced complexity and good accuracy.
Moreover, we propose an SNR region-based enhancement of the approximation method to reduce the complexity even further.
Using Monte-Carlo simulations, the proposed methods are numerically demonstrated for the computation of the mutual information including the entropy term of various channels with finite constellation modulations such as binary and quadratic amplitude modulation (QAM) inputs for communication applications.
\end{abstract}
\begin{IEEEkeywords}
Gaussian mixture distribution, Entropy approximation, Mutual information, Finite input alphabet, Sphere decoding
\end{IEEEkeywords}


\section{Introduction}

\IEEEPARstart{I}{n} general, the computation of Gaussian mixture distributions with a large number of components has a prohibitive computational complexity but a wide range of useful application areas including communications \cite{HBD+08MFI,GVR14TWC,ZSF+03WOC,ALV+06TIT,DOK+14VTC}, data fusion \cite{CWP+11Fusion, HuH08Fusion, ScH09Fusion}, machine learning \cite{CGJ96JAIR, NG08TM}, image and pattern recognition \cite{AK12LNCS,BGP10PR}, and target tracking applications \cite{Sal09TAES,Run07TAES}. 
For instance, the computation of mutual information in communications results in the problem of computing entropy terms of a large system with finite input alphabet which has a prohibitive computational complexity since the number of possible inputs grows exponentially with the system dimension. 
Moreover, in data fusion and target tracking applications, computing the full Gaussian mixture distribution of a sampled data set has prohibitive complexity for high dimensions or a large data set.

In data fusion and tracking areas, Gaussian mixture reduction is common to reduce the problem size and bound the computational complexity and required memory size \cite{CWP+11Fusion, HuH08Fusion, ScH09Fusion, Sal09TAES,Run07TAES}. However, most Gaussian mixture reduction algorithms know the true Gaussian mixture distribution for a sampled data set and start from it to reduce the number of components by merging, pursing, and expanding based on distance measures such as integral squared error (ISE) and Kullback-Leibler (KL) divergence. However, they are intractable for high dimensions since this approach requires the computation of the distance measures among all possible components.
\REV{In this paper, we propose a different approximation approach but in principle, it is also a Gaussian mixture reduction.}

On the other hand, there have been several approaches in communications to approximate the mutual information or the entropy of Gaussian mixture distributions both analytically and numerically.
Huber et al. \cite{HBD+08MFI} proposed an entropy approximation of Gaussian mixture random vectors based on Taylor series expansion, which does not apply to a large system size. 
Girnyk et al. \cite{GVR14TWC} analyzed the capacity of a large multiple input and multiple output (MIMO) system with a finite input alphabet based on the matrix replica method. This approach is only applicable to compute the average capacity of an independent and identically distributed (i.i.d.) MIMO channel with infinite dimension.
Arnold et al. \cite{ALV+06TIT} proposed a simulation-based computation of the mutual information of a time-invariant discrete-time channel with memory. 
Dauwels and Loeliger \cite{DL08TIT} extended the approach to continuous state spaces and Molkaraie and Loeliger \cite{ML13TIT} applied it to information rates computation of two-dimensional channels whose main application is a magnetic recording.
Although this allows the approximation of the mutual information with a long block length, the method is limited to time-invariant frequency-selective fading channels with a relatively short finite impulse response (FIR) length. 
Zhu et al. \cite{ZSF+03WOC} proposed a statistical computation approach for MIMO channels with a finite alphabet depending on the signal-to-noise ratio (SNR).
Even if this approach offers very low complexity for arbitrarily structured channels with high dimension, the accuracy at moderate SNR, especially important for practical systems, is not acceptable. 

In this paper, our main contribution is to provide a novel approximation method with low complexity and good accuracy on the mutual information of arbitrarily structured channels with high dimension, which also leads to new upper and lower bounds.
The main idea is to find $N$-closest Gaussian components through an efficient tree search algorithm and approximate the true Gaussian mixture distribution by a reduced Gaussian mixture distribution.
Based on this approach, we provide upper and lower bounds computable with reduced complexity and, further, an approximation with significantly reduced complexity, which can be computed even for high dimensional cases.
Although we focus on the communication problems in this paper, it is worth mentioning that the proposed method has many general applications where a reduction of the Gaussian mixture is needed.

The rest of this paper is organized as follows. In Section~\ref{sec:system_model}, the problem definition including a basic system model is presented. In Section~\ref{sec:review_SD}, we review the sphere decoding tree search algorithm. Novel sphere decoder approximations on the entropy are provided in Section~\ref{sec:SD_bounds}. In Section~\ref{sec:SD_approximation}, an SNR-based enhanced approximation algorithm suitable for high dimension is proposed. 
In Section~\ref{sec:numerical_examples}, several numerical examples are discussed for various channels. 
Finally, conclusive remarks are provided in Section~\ref{sec:conclusion}.

\section{Problem Definition} \label{sec:system_model}

The Gaussian mixture distribution is a weighted sum of Gaussian distributions with different mean and/or variance, which is mathematically modeled as
\begin{align}
g(\mathbf{x}) = \sum_{i=1}^{N_g} \omega_{i} g_i(\mathbf{x}),
\end{align}
where $\mathbf{x}$ denotes the complex-valued input vector, $N_g$ denotes the total number of Gaussian components, $\omega_i$ denotes the non-negative weight factor for the $i$-th Gaussian component with $\sum_i \omega_i = 1$, and $g_i(\mathbf{x})$ denotes the $i$-th Gaussian component following a complex Gaussian distribution with mean $\mu_i$ and covariance $\mathbf{\Sigma}_i$, i.e.,  $g_i(\mathbf{x})\sim\mathcal{CN}(\mu_i, \mathbf{\Sigma}_i)$. 
For $N_g$ large, the computation of $g(\mathbf{x})$ has a high complexity and therefore reducing the number of components is the main approach of previous Gaussian mixture reduction problem. 

In this paper, we consider the following basic system equation, which is common for many communication systems.
\begin{align}
\mathbf{z} = \mathbf{Hd} + \mathbf{n},
\label{eq:sys_eq}
\end{align}
where $\mathbf{z}\in\mathbb{C}^{N_t\times 1}$ denotes the received signal vector, $\mathbf{d}\in
\mathcal{M}_c^{N_t\times 1}$
denotes the input symbol vector where each symbol $d_k$ is taken from a finite constellation set $\mathcal{M}_c\subset \mathbb{C}$,
$\mathbf{H} \in \mathbb{C}^{N_t \times N_t}$ denotes an arbitrarily structured channel matrix, $\mathbf{n}$ denotes the additive white Gaussian noise vector, 
$\mathbf{n}\sim\mathcal{CN}(0,\mathbf{I})$, and the transmitted power (equivalently, SNR due to normalized unit noise variance) is given $\rho \triangleq \mathbb{E}[\mathbf{d}^{\mathsf{H}}\mathbf{d}]$.
Then, the mutual information between the input $\mathbf{d}$ and the output $\mathbf{z}$ in \eqref{eq:sys_eq} can be expressed by the differential entropies as follows:
\begin{align}
I(\mathbf{z};\mathbf{d}) &= h(\mathbf{z}) - h(\mathbf{z}|\mathbf{d}) = h(\mathbf{z}) - h(\mathbf{n}) \nonumber\\
&= -\mathbb{E}[\log_2(f_{\mathbf{z}}(\mathbf{z}))] - \log_2\left(\det(\pi e \mathbf{I})\right),
\label{eq:mi_prob}
\end{align}
where $f_{\mathbf{z}}(\mathbf{z})$\footnote{We drop the subindex when it is clear from the context.} denotes the probability density function (pdf) of $\mathbf{z}$, which is a Gaussian mixture distribution given by
\begin{align}
f_{\mathbf{z}}(\mathbf{z}) = \sum_{i=1}^{M_c^{N_t}} p(\mathbf{d}_i)f_{\mathbf{z}|\mathbf{d}}(\mathbf{z}|\mathbf{d}_i),
\label{eq:prob_pdf}
\end{align}
where $M_c$ denotes the number of constellation points and $\mathbf{d}_i$ denotes the $i$-th input symbol vector among $M_c^{N_t}$ possibilities. 
For practical communication problems the components of $\mathbf{d}_i$ are usually assumed to be independent and uniformly distributed (i.u.d), i.e., $p(\mathbf{d}_i)=M_c^{-N_t}$.
Note that for large $N_t$, the computation of \eqref{eq:prob_pdf} is infeasible due to the exponentially increasing number of input vectors.
Since the computation of the expectation in \eqref{eq:mi_prob} can be easily handled by Monte-Carlo simulation, the problem at hand is to approximate \eqref{eq:prob_pdf}.
In general, for a given $\mathbf{z}$, only a few terms in the sum in \eqref{eq:prob_pdf} hav a significant contribution. Therefore, finding those components which highly contribute is our main approach for the approximation in the rest of this paper.

\section{A Review of Sphere Decoding Tree Search}\label{sec:review_SD}

Our proposed bounds and approximation presented in next sections are inspired from the sphere decoding (SD) algorithm \cite{FiP85MC,AEV+02TIT,DGC03TIT,MGD+06TIT,HaV05TSP,ViH05TSP,JaO05TSP,GuN06JSAC,BaT08TWC}, which is a well-known maximum likelihood (ML) branch and bound algorithm in a tree search for MIMO detection, i.e., finding the most likely input vector $\mathbf{d}_i$ in given the received vector $\mathbf{z}$, and the soft SD algorithm \cite{BGB+03GC} which principle can be used for capacity approximation as shown in the following. The motivation is that it can reduce the search space and, thus, the required computations via an efficient tree search. Here, we briefly review the SD algorithm.

\begin{figure}[tb]
  \centering
  \includegraphics[width=3.5in]{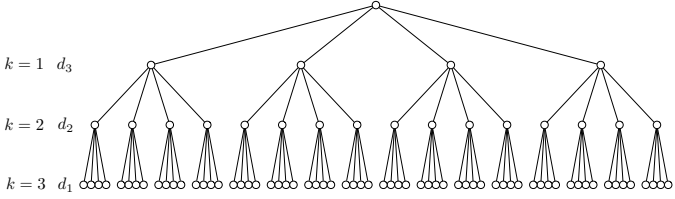}
  \caption{An example of SD search tree (e.g., 4-QAM and $N_t=3$).}
  \label{fig:tree_example}
\end{figure}

In order to construct a search tree, the SD algorithm first performs QR factorization of the channel matrix $\mathbf{H}$. Then, the system equation \eqref{eq:sys_eq} is equivalently given by
\begin{align}
\mathbf{v} = \mathbf{Rd} + \mathbf{w},
\label{eq:sys_eq_qr}
\end{align}
with $\mathbf{H}=\mathbf{QR}$ in which $\mathbf{Q}$ is a unitary matrix and $\mathbf{R}$ is an upper triangular matrix, $\mathbf{v}=\mathbf{Q}^{\mathsf{H}}\mathbf{z}$, $\mathbf{w}=\mathbf{Q}^{\mathsf{H}}\mathbf{n}\sim\mathcal{CN}(0,\mathbf{I})$, and $\mathbf{d}=[d_1,\ldots,d_{N_t}]^{\mathsf{T}}$. 
It is worth noting that since any invertible linear operation does not change the mutual information \cite{Gal68Wil}, $I(\mathbf{z};\mathbf{d})=I(\mathbf{v};\mathbf{d})$.
Then, a search tree is constructed from the bottom to the top of the equivalent upper-triangular channel matrix $\mathbf{R}$. That is, first branches from the root node are constructed from the last diagonal term of $\mathbf{R}$ corresponding to $d_{N_t}$ until the last branches to the leaf nodes are constructed from the first row of $\mathbf{R}$ corresponding to $d_1$. Let $r_{ij}$ denote the $(i,j)$-th element of $\mathbf{R}$. Then, at the $k$-th depth, the cost value corresponding to the Euclidean distance between the received vector $\mathbf{v}$ and the considered input $\mathbf{d}$ can be recursively expressed as
\begin{align}\label{eq:c_k}
c(k,\mathbf{d}_{N_t-k+1}^{N_t}) &= c(k-1,\mathbf{d}_{N_t - k + 2}^{N_t}) \nonumber\\
&\quad+ \Big| v_{N_t-k+1} - \sum_{j=N_t-k+2}^{N_t} r_{k,j} d_{j}  \Big|^2
\end{align}
where $k\in\{1,\ldots,N_t\}$, $c(0,\mathbf{d}_{N_t +1}^{N_t})=0$, $\mathbf{d}_{i}^{j} \triangleq [d_{i}, d_{i+1}, \ldots, d_{j}]^{\mathsf{T}}$, and $\mathbf{v}=[v_1,\ldots,v_{N_t}]^{\mathsf{T}}$.
Fig.~\ref{fig:tree_example} illustrates an example of SD search tree construction for case of 4-quadratic amplitude modulation (QAM) and $N_t=3$ resulting in $4^3=64$ possibilities.

\subsection{Depth-First Search (DFS)}

\begin{figure}[tb]
  \centering
  \subfigure[]{\includegraphics[width=6.5cm]{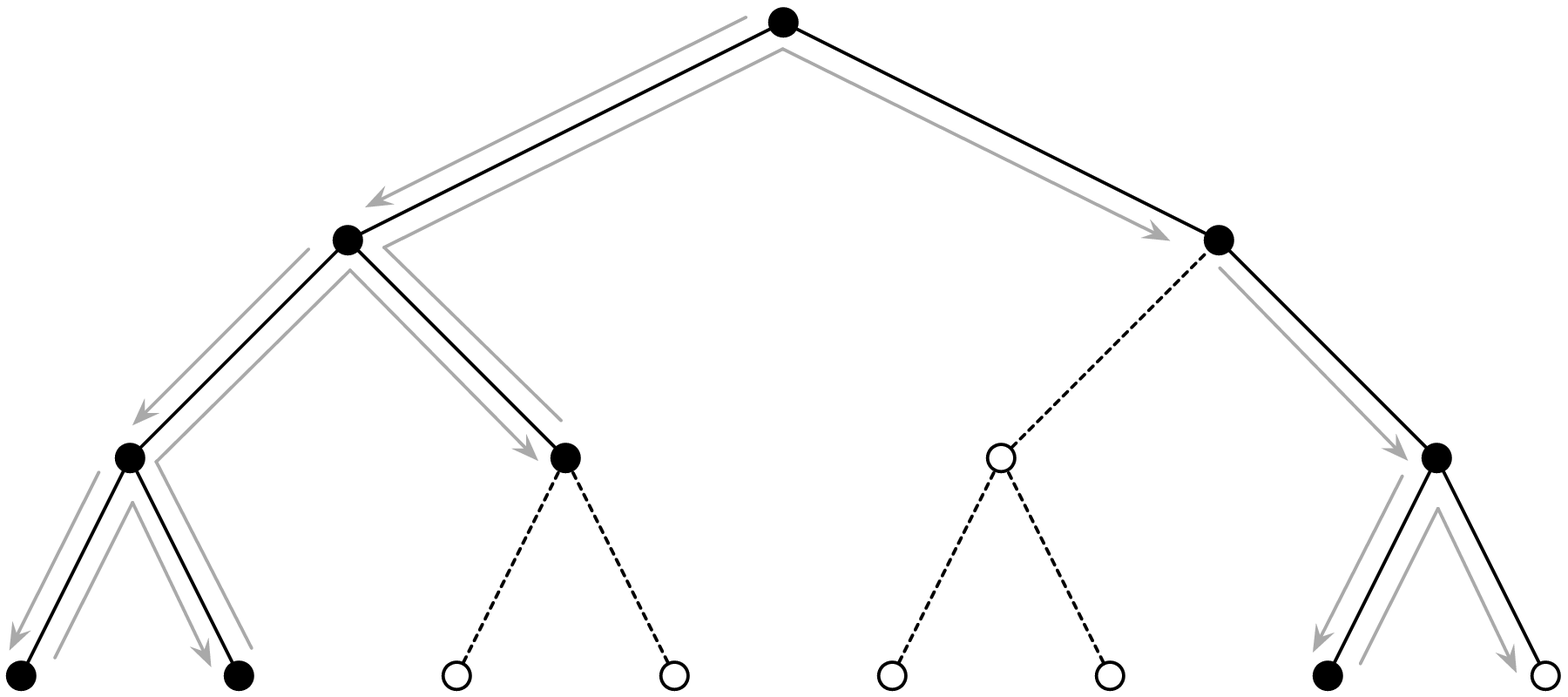}}
  \subfigure[]{\includegraphics[width=6.5cm]{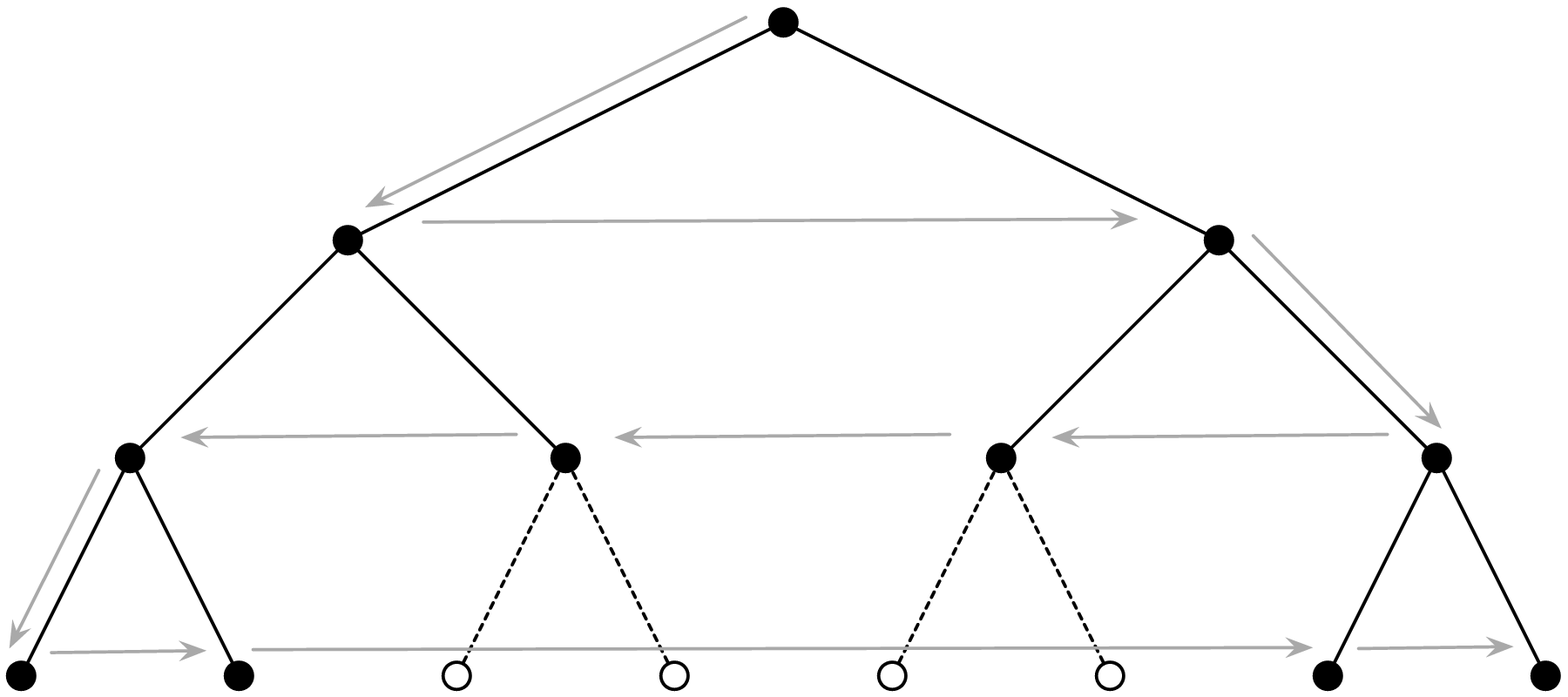}}
  \caption{Examples of (a) DFS and (b) BFS (e.g., binary input and $N_t=3$). The gray arrows denote the search movements. The black/white circle denotes the visited/non-visited node. The dashed line denotes the pruned branch.}
\label{fig:DFS_BFS}
\end{figure}

The DFS algorithm searches for components with the distance less than the sphere radius in both forward and backward directions among the sub-trees.
It first goes through the search tree by a leaf node in the forward direction of $k=1,2,\ldots, N_t$ and then it moves backward in the direction of $N_t,N_t-1,\ldots,1$.
Fig.~\ref{fig:DFS_BFS}~(a) illustrates an example of the DFS.

The DFS algorithm efficiently provides the optimal ML solution corresponding to the closest input symbol vector for traditional MIMO detection. Moreover, during the tree search, if it finds an input symbol vector with shorter distance than the sphere radius, the sphere radius can be dynamically updated which reduces the tree search complexity for the purpose of finding only the closest component.
However, in this paper, our purpose of the tree search is finding all components within a given sphere radius. Therefore, we use a fixed sphere radius and do not consider its dynamic update.
As a result, after the tree search, it is guaranteed to find all input symbol vectors with shorter distance than the sphere radius. 
Denoting the number of components within the sphere radius as $N$, the $N$-closest components\footnote{We can also fix the number of components $N$ and update the sphere radius as often as $N$ components are found. Then, we have $N$ candidates found during the tree search.} can be found during the tree search.

\subsection{Breadth-First Search (BFS)}

The BFS algorithm searches for components in the forward direction only. That is, it searches all nodes at a certain depth and then moves to the next depth.
Fig.~\ref{fig:DFS_BFS}~(b) illustrates an example of the BFS.

In most applications of MIMO detection, the BFS algorithm keeps just $K$-best components and prune the other branches at each depth. This is called $K$-Best SD algorithm \cite{WTC+02ISCS,RBO04ICC,GuN06JSAC}.
In this case, if $K$ is sufficiently large, the solution approaches the optimal ML solution. 
In contrast, limiting $K$ reduces the search complexity and thus it provides a fixed search complexity.
This is the main advantage of the $K$-best SD algorithm since it is easily implemented in a parallel and a pipelined fashion.
In the viewpoint of finding $N$-closest components in our problem, this approach also can provide the fixed complexity relying on $K$ even though the components found at the end are not guaranteed to be the $N$-closest components.

\section{Sphere Decoder Approximation}\label{sec:SD_bounds}

In this section, we exploit the SD algorithm in a different manner in order to find approximations and bounds on the entropy of Gaussian mixture distributions.
While the aim of original SD algorithm is to find only the closest input vector, we find the $N$-closest input vectors, which contribute the most to $f(\mathbf{z})$, through an efficient tree search.
We propose two approaches employing both the DFS and the BFS. The two approaches give different accuracy and complexity control methods although the basic principle is the same.
The following bounds are the approximation. From the simulations, we see that the upper bound is usually close to the true curve (refer to Fig.~\ref{fig:CH10_all}~(a), Fig.~\ref{fig:CH10_up_lo}, and Fig.~\ref{fig:Nt8}~(a)).

\subsection{DFS-Based Upper and Lower Bounds}

Starting from \eqref{eq:sys_eq_qr}, the DFS-based algorithm finds input symbol vectors satisfying
\begin{align}
\|\mathbf{v} - \mathbf{Rd}\|^2 \leq \zeta^2,
\end{align}
where the sphere radius is set to
\begin{align}
\zeta^2=\alpha\|\mathbf{v} - \mathbf{Rd}_{0}  \|^2,
\label{eq:sphere_radius}
\end{align}
where $\mathbf{d}_{0}$ denotes the Babai estimate\footnote{Equivalently, it is the zero-forcing (ZF) point found as $\mathbf{d}_{0}=\mathbf{H}^{\dagger}\mathbf{z}$ where $\mathbf{H}^{\dagger} = (\mathbf{H}^{\mathsf{H}}\mathbf{H})^{-1}\mathbf{H}^{\mathsf{H}}$.} \cite{Babai} and $\alpha$ denotes a control parameter which can be used to adjust complexity versus accuracy. \REV{If we increase $\alpha$, the accuracy increases since the search result can include more components due to the larger search radius, while the complexity also increases since it requires more searches in the tree. It gives the full tree search when $\alpha\rightarrow\infty$, i.e., the true distribution. Note that if $\alpha \geq 1$, the sphere radius \eqref{eq:sphere_radius} guarantees to find at least one component in the tree search because it includes at least $\mathbf{d}_0$.}
After the SD tree search, the following set of ordered symbol vectors are found:
\begin{align}
\mathcal{D}_{\mathrm{DFS}}^{(\zeta)}=\{\hat{\mathbf{d}}_1, \hat{\mathbf{d}}_2, \ldots, \hat{\mathbf{d}}_{N_{\mathrm{DFS}}^{(\zeta)}}\},
\end{align}
where $\mathcal{D}_{\mathrm{DFS}}^{(\zeta)} \subset \mathcal{D} = \mathcal{D}_{\mathrm{DFS}}^{(\infty)}$, $|\mathcal{D}|=M_c^{N_t}$, $N_{\mathrm{DFS}}^{(\zeta)} = |\mathcal{D}_{\mathrm{DFS}}^{(\zeta)}|$, and $\|\mathbf{v} - \mathbf{R}\hat{\mathbf{d}}_1\|^2 \leq \|\mathbf{v} - \mathbf{R}\hat{\mathbf{d}}_2\|^2 \leq \ldots \leq \|\mathbf{v} - \mathbf{R}\hat{\mathbf{d}}_{N_{\mathrm{DFS}}^{(\zeta)}}\|^2$. 
Assuming i.u.d. input $\hat{\mathbf{d}}$, the true pdf $f(\mathbf{z})$ can be expressed as:
\begin{align}
f_{\mathbf{z}}(\mathbf{z}) &= \sum_{\hat{\mathbf{d}}\in\mathcal{D}} p(\hat{\mathbf{d}}) f_{\mathbf{z}|\mathbf{d}}(\mathbf{z}|\hat{\mathbf{d}})
= \sum_{\hat{\mathbf{d}}\in\mathcal{D}} p(\hat{\mathbf{d}}) f_{\mathbf{v}|\mathbf{d}}(\mathbf{v}|\hat{\mathbf{d}}) \nonumber\\
&=   \frac{1}{M_c^{N_t}} \cdot \frac{1}{\pi^{N_t}} \sum_{\hat{\mathbf{d}}\in\mathcal{D}}\exp\left(-{\|\mathbf{v}-\mathbf{R}\hat{\mathbf{d}}\|^2}\right), \label{eq:pdf_true}
\end{align}
where the second equality is obtained from the fact that $\|\mathbf{v}-\mathbf{R}\hat{\mathbf{d}}\|^2 = \|\mathbf{z}-\mathbf{H}\hat{\mathbf{d}}\|^2$ due to unitary $\mathbf{Q}$.
Therefore, $f_{\mathbf{z}}(\mathbf{z})$ is equal to $f_{\mathbf{v}}(\mathbf{v})$. Accordingly, we have $h(\mathbf{z})=h(\mathbf{v})$ and $I(\mathbf{z};\mathbf{d})=I(\mathbf{v};\mathbf{d})$.
Let ${T} \triangleq  \sum_{\hat{\mathbf{d}}\in\mathcal{D}} \exp\left(-D(\mathbf{\hat{\mathbf{d}}})\right)$ in \eqref{eq:pdf_true} where  $D(\mathbf{x})\triangleq \|\mathbf{v}-\mathbf{R}\mathbf{x}\|^2$.
For the ordered input symbol vectors with respect to the distance, i.e., $\mathcal{D}=\{\hat{\mathbf{d}}_1, \hat{\mathbf{d}}_2, \ldots, \hat{\mathbf{d}}_{N_t}\}$, the following relations hold after the SD tree search:
\begin{align}
&\exp(-D(\hat{\mathbf{d}}_{1})) \geq \cdots \geq \exp(-D(\hat{\mathbf{d}}_{N_{\mathrm{DFS}}^{(\zeta)}})) \geq \exp(-\zeta^2) \nonumber\\
&> \exp(-D(\hat{\mathbf{d}}_{N_{\mathrm{DFS}}^{(\zeta)}+1})) \geq \cdots \geq \exp(-D(\hat{\mathbf{d}}_{N_t})).
\label{eq:exp_relation}
\end{align}
Thus, $T$ can be expressed in two parts:
\begin{align}
T &= \underbrace{ \sum_{\hat{\mathbf{d}} \in \mathcal{D}_{\mathrm{DFS}}^{(\zeta)}} \exp(-D(\hat{\mathbf{d}})) }_{\text{components found}} + \underbrace{ \sum_{\hat{\mathbf{d}} \in \mathcal{D} \backslash \mathcal{D}_{\mathrm{DFS}}^{(\zeta)}}  \exp(-D(\hat{\mathbf{d}})) }_{\text{components pruned}}.
\label{eq:T_DFS}
\end{align}
The second term for pruned components is upper-bounded by $\sum_{k=1}^{|\mathcal{D}|-N_{\mathrm{DFS}}^{(\zeta)}}\exp(-\zeta^2)$.
Therefore, ${T}$ can be bounded as follows
\begin{align}
&\sum_{\hat{\mathbf{d}}\in\mathcal{D}_{\mathrm{DFS}}^{(\zeta)}} \exp\left( - \|\mathbf{v}-\mathbf{R}\hat{\mathbf{d}}\|^2 \right) \leq {T} \nonumber\\
& < \sum_{\hat{\mathbf{d}}\in\mathcal{D}_{\mathrm{DFS}}^{(\zeta)}} \exp\left( - \|\mathbf{v}-\mathbf{R}\hat{\mathbf{d}}\|^2 \right) + \left(|\mathcal{D}| - N_{\mathrm{DFS}}^{(\zeta)}\right) \exp\left(-\zeta^2\right). 
\label{eq:bounds_T}
\end{align}
Let us define $\underline{f}_{\mathrm{DFS}}(\mathbf{v})$ and $\overline{f}_{\mathrm{DFS}}(\mathbf{v})$ by
\begin{align}
\underline{f}_{\mathrm{DFS}}(\mathbf{v}) &\triangleq\sum_{\hat{\mathbf{d}}\in\mathcal{D}_{\mathrm{DFS}}^{(\zeta)}} \frac{1}{(\pi M_c)^{N_t}}\exp\left( - \|\mathbf{v}-\mathbf{R}\hat{\mathbf{d}}\|^2 \right), \label{eq:f_v_hi_DFS}\\
\overline{f}_{\mathrm{DFS}}(\mathbf{v}) &\triangleq \underline{f}_{\mathrm{DFS}}(\mathbf{v}) + \frac{|\mathcal{D}| - N_{\mathrm{DFS}}^{(\zeta)}}{(\pi M_c)^{N_t}} \exp\left(-\zeta^2\right). \label{eq:f_v_lo_DFS}
\end{align}
Then, the differential entropy of $\mathbf{z}$ is bounded by
\begin{align}
{h}^{lo}_{\mathrm{DFS}} < h(\mathbf{z}) \leq h_{\mathrm{DFS}}^{up},
\label{eq:h_z_bounds_DFS}
\end{align}
where $h_{\mathrm{DFS}}^{lo} = -\mathbb{E}\left[ \log_2 \overline{f}_{\mathrm{DFS}}(\mathbf{v})\right]$ and $h_{\mathrm{DFS}}^{up} = -\mathbb{E} \left[ \log_2 \underline{f}_{\mathrm{DFS}}(\mathbf{v})\right]$ 
since $\underline{f}_{\mathrm{DFS}}(\mathbf{v}) \leq f(\mathbf{z}) < \overline{f}_{\mathrm{DFS}}(\mathbf{v})$ for all $\mathbf{v}=\mathbf{Q}^{\mathsf{H}}\mathbf{z}$.\\

\subsubsection*{Enhanced Lower Bound}
During the tree search, a pruned branch including sub-branches has a distance value greater than $\zeta^2$. 
Let the cost value of the pruned branch at the $k$-th depth of the search tree be denoted by $c(k, \mathbf{d}_{N_t-k+1}^{N_t})$ where $\mathbf{d}_{N_t-k+1}^{N_t} = [d_{N_t-k+1}, \ldots, d_{N_t}]^{\mathsf{T}}$ is the input symbol vector with length $k$ found in previous and current depth searches. Then, the pruned branch includes $M_c^{N_t-k}$ sub-branches and the symbol vectors corresponding to the sub-branches can use $c(k, \mathbf{d}_{N_t-k+1}^{N_t})$ instead of $\zeta^2$ for the \REV{$\exp(-\zeta^2)$ term} in \eqref{eq:f_v_lo_DFS}. 

In more detail, denote the remaining Euclidean distance values at leaf nodes for each sub-branch by $\bar{c}(N_t, \mathbf{d}_{1}^{N_t - k}) \triangleq c(N_t, \mathbf{d}_{1}^{N_t}) - c(k, \mathbf{d}_{N_t-k+1}^{N_t}) \geq 0$ where $\mathbf{d}_{i}^{j} = [d_{i}, \ldots, d_{j}]^{\mathsf{T}}$.  Since for the pruned branch, $\zeta^2 < c(k, \mathbf{d}_{N_t-k+1}^{N_t}) \leq c(N_t, \mathbf{d}_{1}^{N_t}) = c(k, \mathbf{d}_{N_t-k+1}^{N_t}) + \bar{c}(N_t, \mathbf{d}_{1}^{N_t-k})$, replacing $\zeta^2$ by $c(k, \mathbf{d}_{N_t-k+1}^{N_t})$ for all the pruned branches yields a better lower bound on the entropy.

Let us define $\overline{f}^{+}_{\mathrm{DFS}}(\mathbf{v})$ by
\begin{align}
\overline{f}^{+}_{\mathrm{DFS}}(\mathbf{v}) &\triangleq \underline{f}_{\mathrm{DFS}}(\mathbf{v}) + \frac{1}{(\pi M_c)^{N_t}}\sum_{\hat{\mathbf{d}} \in \mathcal{D} \backslash \mathcal{D}_{\mathrm{DFS}}^{(\zeta)}}  \exp(-\tilde{c}(\hat{\mathbf{d}})),
\end{align}
where $\tilde{c}(\hat{\mathbf{d}})$ denotes the cost value of $\hat{\mathbf{d}}$ at its own pruned depth. For instance, if $\hat{\mathbf{d}}$ is pruned at depth $k$, $\tilde{c}(\hat{\mathbf{d}}) = c(k, \hat{\mathbf{d}}_{N_t-k+1}^{N_t})$. 
Then, the differential entropy of $\mathbf{z}$ gets the enhanced lower bound as
\begin{align}
{h}^{lo}_{\mathrm{DFS}} < {h}^{lo+}_{\mathrm{DFS}} < h(\mathbf{z}).
\label{eq:h_z_lo+_bound_DFS}
\end{align}
where ${h}^{lo+}_{\mathrm{DFS}}=-\mathbb{E}\left[ \log_2 \overline{f}^{+}_{\mathrm{DFS}}(\mathbf{v})\right]$.
Substituting the entropy bounds into \eqref{eq:mi_prob} results in bounds as follows:
\begin{align}
I_{\mathrm{DFS}}^{lo} < I_{\mathrm{DFS}}^{lo+} < I(\mathbf{z};\mathbf{d}) \leq I_{\mathrm{DFS}}^{up}.
\label{eq:MI_bounds_DFS}
\end{align}

\subsection{BFS-Based Upper and Lower Bounds}

For BFS-based upper and lower bounds, we employ BFS-based $K$-best SD approach.
Similarly to the DFS-based algorithm, the BFS-based algorithm finds input symbol vectors satisfying 
\begin{align}
\|\mathbf{v} - \mathbf{Rd}\|^2 \leq \zeta^2, \nonumber
\end{align}
but $\zeta^2$ is set to a sufficiently large value so that all components are included within the sphere radius.
Differently from the DFS-based algorithm, the BFS-based algorithm finds the $K$-closest components at each depth (i.e., each breadth). 
In more detail, it takes $K$ shortest distance components among $M_c K$ components at each $k$-th depth. Note that when $M_c^k < K$, all $M_c^k$ components are taken at the depth.
After all, $K$ becomes a control parameter in the BFS-based algorithm to adjust complexity versus accuracy instead of the $\alpha$ parameter in the DFS-based algorithm.
Note that if $K \geq M_c^{N_t-1}$, all the components are found at the end of the tree search in the BFS-based algorithm.

After the SD tree search, the following set of ordered symbol vectors are found:
\begin{align}
\mathcal{D}_{\mathrm{BFS}}^{(K)}=\{\hat{\mathbf{d}}_1, \hat{\mathbf{d}}_2, \ldots, \hat{\mathbf{d}}_{N_{\mathrm{BFS}}^{(K)}}\},
\end{align}
where $\mathcal{D}_{\mathrm{BFS}}^{(K)} \subset \mathcal{D} = \mathcal{D}_{\mathrm{BFS}}^{(\infty)}$, $|\mathcal{D}|=M_c^{N_t}$, $N_{\mathrm{BFS}}^{(K)} = |\mathcal{D}_{\mathrm{BFS}}^{(K)}|$, and $\|\mathbf{v} - \mathbf{R}\hat{\mathbf{d}}_1\|^2 \leq \|\mathbf{v} - \mathbf{R}\hat{\mathbf{d}}_2\|^2 \leq \ldots \leq \|\mathbf{v} - \mathbf{R}\hat{\mathbf{d}}_{N_{\mathrm{BFS}}^{(K)}}\|^2$. 

In the BFS-based algorithm, the corresponding relation to \eqref{eq:exp_relation} does not hold since the components found are not exactly the $N$-closest components anymore.
However, \eqref{eq:T_DFS} can be still equivalently expressed as
\begin{align}
T &= \underbrace{ \sum_{\hat{\mathbf{d}} \in \mathcal{D}_{\mathrm{BFS}}^{(K)}} \exp(-D(\hat{\mathbf{d}})) }_{\text{components found}} + \underbrace{ \sum_{\hat{\mathbf{d}} \in \mathcal{D} \backslash \mathcal{D}_{\mathrm{BFS}}^{(K)}}   \exp(-D(\hat{\mathbf{d}})) }_{\text{components pruned}}.
\label{eq:T_BFS}
\end{align}
Thus, $T$ is lower-bounded by the first term of the right-hand side of \eqref{eq:T_BFS}.
Although we cannot find an upper bound as in \eqref{eq:bounds_T}, the enhanced lower bound approach on the entropy still works in this case.

Let us define $\underline{f}_{\mathrm{BFS}}(\mathbf{v})$ and $\overline{f}^{+}_{\mathrm{BFS}}(\mathbf{v})$ by
\begin{align}
\underline{f}_{\mathrm{BFS}}(\mathbf{v}) &\triangleq\sum_{\hat{\mathbf{d}}\in\mathcal{D}_{\mathrm{BFS}}^{(\zeta)}} \frac{1}{(\pi M_c)^{N_t}}\exp\left( - \|\mathbf{v}-\mathbf{R}\hat{\mathbf{d}}\|^2 \right), \label{eq:f_v_hi_BFS}\\
\overline{f}^{+}_{\mathrm{BFS}}(\mathbf{v}) &\triangleq \underline{f}_{\mathrm{BFS}}(\mathbf{v}) + \frac{1}{(\pi M_c)^{N_t}}\sum_{\hat{\mathbf{d}} \in \mathcal{D} \backslash \mathcal{D}_{\mathrm{BFS}}^{(K)}}  \exp(-\tilde{c}(\hat{\mathbf{d}})), \label{eq:f_v_lo+_BFS}
\end{align}
where $\tilde{c}(\hat{\mathbf{d}})$ denotes the cost value of $\hat{\mathbf{d}}$ at its own pruned depth.
Then, the differential entropy of $\mathbf{z}$ is bounded by
\begin{align}
\hat{h}^{lo+}_{\mathrm{BFS}} < h(\mathbf{z}) \leq \hat{h}^{up}_{\mathrm{BFS}},
\label{eq:h_z_bounds_BFS}
\end{align}
where $\hat{h}^{lo+}_{\mathrm{BFS}}=-\mathbb{E}\left[ \log_2 \overline{f}^{+}_{\mathrm{BFS}}(\mathbf{v})\right]$ and $\hat{h}^{up}_{\mathrm{BFS}}=-\mathbb{E} \left[ \log_2 \underline{f}_{\mathrm{BFS}}(\mathbf{v})\right]$ since $\underline{f}_{\mathrm{BFS}}(\mathbf{v}) \leq f(\mathbf{z}) < \overline{f}^{+}_{\mathrm{BFS}}(\mathbf{v})$.
Substituting the entropy bounds into \eqref{eq:mi_prob} results in bounds as follows:
\begin{align}
I_{\mathrm{BFS}}^{lo+} < I(\mathbf{z};\mathbf{d}) \leq I_{\mathrm{BFS}}^{up}.
\label{eq:MI_bounds_DFS}
\end{align}

\subsubsection*{Determination of the $K$ Parameter}
The BFS-based bounds algorithm enables the complexity\footnote{Throughout this paper, the complexity is evaluated in terms of the number of visited nodes in a tree search, which is common in the literature on the sphere decoding algorithms \cite{HaV05TSP,JaO05TSP}.} to be fixed as a certain value by adjusting $K$ parameter, while the DFS-based bounds algorithm can implicitly control the complexity according to $\alpha$ parameter. 
Define $k_0 \triangleq \max\left\{k: M_c^{k-1} < K\right\}$.  
Then, the complexity of the bounds based on the BFS algorithm in terms of the number of visited nodes in the tree search is given by
\begin{align}
\mathcal{C}(K) & = \sum_{k=1}^{k_0} M_c^k + \sum_{k=k_0+1}^{N_t} M_c K \nonumber\\
& = \frac{M_c (1 - M_c^{k_0})}{1 - M_c} + \left( N_t - k_0 \right) M_c K.
\label{eq:C_K}
\end{align}
Note that for $K \rightarrow \infty$, we have $\mathcal{C}(\infty) = \sum_{k=1}^{N_t} M_c^k = \frac{M_c (1 - M_c^{N_t})}{1 - M_c}$, which is the complexity of the true Gaussian mixture distribution.
Finally, for a given complexity $\mathcal{C}_0$, the $K$ parameter is determined by
\begin{align}\label{eq:K_C0}
K(\mathcal{C}_0) = \left\lfloor\frac{1}{N_t - k_0} \left( \frac{\mathcal{C}_0}{M_c} - \frac{M_c^{k_0} - 1}{M_c - 1} \right)\right\rfloor.
\end{align}

Table~\ref{tbl:param} illustrates the notations used in algorithm descriptions in the following.
The overall procedure of the proposed SD approximation algorithm is specified in Algorithm~\ref{alg:SD_bound}.
The DFS-based and BFS-based SD tree search algorithms used in Algorithm~~\ref{alg:SD_bound} are described as recursive functions in Algorithm~\ref{alg:DFS} and Algorithm~\ref{alg:BFS}, respectively.

\begin{table}[tb]
\caption{Notations used in algorithms}
\label{tbl:param}
\centering
\begin{tabular}{|c|c|}
\hline 
Notation & Description \\ 
\hline 
$N_d$ & Number of iterations for generating $\mathbf{d}$ \\ 
\hline 
$N_n$ & Number of iterations for generating $\mathbf{n}$ \\ 
\hline 
$\mathcal{U}(\mathcal{M}^{N_t})$ & Uniform distribution on the $N_t$-dimension Cartesian\\
& product of the constellation points set $\mathcal{M}$\\ 
\hline 
$\hat{h}^{\mathrm{y}}_{\mathrm{x}}$ & Monte-Carlo integration approximation for the entropy\\ 
\hline
\end{tabular} 
\end{table}

\begin{algorithm}\label{alg:SD_bound}
\DontPrintSemicolon
   \caption{Sphere Decoder Approximation}
   \KwIn{$\mathbf{H},\rho$}
   \KwOut{$\hat{h}^{up}_{\mathrm{SD}},\hat{h}^{lo}_{\mathrm{SD}},\hat{h}^{lo+}_{\mathrm{SD}}$}
   \BlankLine
   $[\mathbf{Q}~\mathbf{R}]\gets \mathrm{qr}({\mathbf{H}})$\tcp*[r]{QR factorization}
   \tcp*[l]{Integration by a Monte-Carlo method}
   \For(\tcp*[f]{Loop for $\mathbf{d}$}){$i = 1$ \KwTo $N_{d}$}
   {
      Generate $\mathbf{d}^{(i)} \gets \sqrt{\rho}\cdot \mathbf{s}$ where $\mathbf{s}\sim\mathcal{U}(\mathcal{M}^{N_t})$\;
      \For(\tcp*[f]{Loop for $\mathbf{n}$}){$j = 1$ \KwTo $N_{n}$}
      {
      	Generate $\mathbf{n}^{(j)}$ where $\mathbf{n}^{(j)}\sim\mathcal{CN}(0,\mathbf{I})$\;
		$\mathbf{z}^{(i,j)} \gets {\mathbf{H}}\mathbf{d}^{(i)} + \mathbf{n}^{(j)}$\;
		$\mathbf{v}^{(i,j)} \gets \mathbf{Q}^{\mathsf{H}}\mathbf{z}^{(i,j)}$\;
		\tcp*[l]{Babai estimate}
		${\mathbf{d}}_{0}^{(j)} \gets ({\mathbf{H}}^{\mathsf{H}}{\mathbf{H}})^{-1}{\mathbf{H}}^{\mathsf{H}} \mathbf{z}^{(i,j)}$\;
		\tcp*[l]{Call a tree search algorithm}    	
		\uIf{$\mathrm{DFS}$}{
			Set $\alpha\geq 1$ and		
			$\zeta^{2} \gets \alpha \| \mathbf{v}^{(i,j)} -  \mathbf{R}{\mathbf{d}}_{0}^{(j)} \|^2$\;  		
			$[\mathcal{D}_{\mathrm{SD}}, \mathcal{E}]\gets$~\DFS{$\{\mathbf{v}^{(i,j)},\mathbf{R},\zeta^2\}$,$\{1,[~],0,0,\emptyset\}$}\;
		}  
		\ElseIf{$\mathrm{BFS}$}{
			Set $K$ according to \eqref{eq:K_C0}\;  	
			$[\mathcal{D}_{\mathrm{SD}}, \mathcal{E}]\gets$~\BFS{$\{\mathbf{v}^{(i,j)},\mathbf{R},K\}$,$\{1,\emptyset,\emptyset,0\}$}\;			
		}  
		\tcp*[l]{Compute pdfs}
		$\underline{f}^{(i,j)} \gets \sum_{\hat{\mathbf{d}}\in\mathcal{D}_{\mathrm{SD}}} \frac{1}{M_c^{N_t}}\exp(-\|\mathbf{v}^{(i,j)}-{\mathbf{H}}\hat{\mathbf{d}}\|^2)$\;		
		$\overline{f}^{(i,j)} \gets \underline{f}^{(i,j)} + \frac{|\mathcal{D}| - |\mathcal{D}_{\mathrm{SD}}|}{(M_c \cdot \pi)^{N_t}} \exp\left(-\zeta^2\right)$\;
		$\overline{f}^{+(i,j)} \gets \underline{f}^{(i,j)} + \mathcal{E}$\; 
      }
      \tcp*[l]{Compute entropy bounds}
      $\hat{h}^{up}_{\mathrm{SD}}\gets -\frac{1}{N_d N_n}\sum_{i=1}^{N_d}\sum_{j=1}^{N_n}\log_2\left(\underline{f}^{(i,j)}\right)$\;
	  $\hat{h}^{lo}_{\mathrm{SD}}\gets -\frac{1}{N_d N_n}\sum_{i=1}^{N_d}\sum_{j=1}^{N_n}\log_2\left(\overline{f}^{(i,j)}\right)$\;
	  $\hat{h}^{lo+}_{\mathrm{SD}}\gets -\frac{1}{N_d N_n}\sum_{i=1}^{N_d}\sum_{j=1}^{N_n}\log_2\left(\overline{f}^{+(i,j)}\right)$\;
   }
\end{algorithm}

\begin{algorithm}\label{alg:DFS}
\SetKwFunction{FnDFS}{\textbf{Function}~\DFS}
\DontPrintSemicolon
\caption{DFS-Based SD Tree Search}     
   \FnDFS{$\{\mathbf{v},\mathbf{R},\zeta^2\},\{k,\mathbf{d},c,\mathcal{E},{\mathcal{D}}_{\mathrm{DFS}}^{(\zeta)}\}$}{\;
   Store $\mathbf{d}^{\prime} \gets \mathbf{d}$ and $c^{\prime} \gets c$\;
   \For{$m\gets 1$~\KwTo~$M_c$}{
   		$\mathbf{d} \gets[d_m;\mathbf{d}^{\prime}]$ where $d_m\gets\mathcal{M}(m)$\;
   		Compute the cost value $c$ according to \eqref{eq:c_k}\;
   		\If(\tcp*[f]{Valid: Searching}){$c \leq \zeta^2$}{
   			\If(\tcp*[f]{Leaf node}){$k=N_t$}{
   				${\mathcal{D}}_{\mathrm{DFS}}^{(\zeta)} \gets {\mathcal{D}}_{\mathrm{DFS}}^{(\zeta)}\bigcup\{\mathbf{d}\}$\;
   			}
   			\Else(\tcp*[f]{Intermediate node}){
   				\tcp*[l]{Go to next depth}
   				\DFS{$\{\mathbf{v},\mathbf{R},\zeta^2\},\{k+1,\mathbf{d},c,\mathcal{E},{\mathcal{D}}_{\mathrm{DFS}}^{(\zeta)}\}$}\;
   			}
   		}
   		\Else(\tcp*[f]{Invalid: Pruning}){
   			\tcp*[l]{Update the exponential term for enhanced lower bound}
   			$\mathcal{E} \gets \mathcal{E} + \exp(-c)\cdot M_c^{N_t-k}$\;
   		}
   }
   \KwRet{$\mathcal{D}_{\mathrm{DFS}}^{(\zeta)}, \mathcal{E}$}
   }
\end{algorithm}

\begin{algorithm}\label{alg:BFS}
\SetKwFunction{FnBFS}{\textbf{Function}~\BFS}
\DontPrintSemicolon
\caption{BFS-Based SD Tree Search}   
\FnBFS{$\{\mathbf{v},\mathbf{R},K\},\{k,\mathcal{D},\mathcal{C},\mathcal{E}\}$}{\;
   Set $\mathcal{D}_{\mathrm{cand}}\gets\emptyset$ and $\mathcal{C}_{\mathrm{cand}}\gets\emptyset$\;
   $K^{\prime}\gets\min\{K,M_c^{k-1}\}$\tcp*[r]{For $K>M_c^{k-1}$}   
   \For{$i=1$~\KwTo~$K^{\prime}$}{
   		$\mathbf{d}^{\prime} \gets \mathcal{D}(i)$ and $c^{\prime} \gets \mathcal{C}(i)$\tcp*[r]{$i$-th element}
   		\For{$m=1$~\KwTo~$M_c$}{
   			$\mathbf{d} \gets[d_m;\mathbf{d}^{\prime}]$ where $d_m\gets\mathcal{M}(m)$\;
   			Compute the cost value $c$ according to \eqref{eq:c_k}\;
   			$\mathcal{D}_{\mathrm{cand}}\gets\mathcal{D}_{\mathrm{cand}}\bigcup\{\mathbf{d}\}$\;
   			$\mathcal{C}_{\mathrm{cand}}\gets \mathcal{C}_{\mathrm{cand}} \bigcup \{c\}$\;
   		}   
   }
   \tcp*[l]{Sort based on the cost values}
   $[\mathcal{D}_{\mathrm{sort}},\mathcal{C}_{\mathrm{sort}}]\gets$~\sort{$\mathcal{D}_{\mathrm{cand}},\mathcal{C}_{\mathrm{cand}}$}\;
   \If(\tcp*[f]{Leaf node}){$k=N_t$}{
   		$\mathcal{D}_{\mathrm{BFS}}^{(K)}\gets \mathcal{D}_{\mathrm{sort}}$\;
   }
   \Else(\tcp*[f]{Intermediate node}){
   		\tcp*[l]{Take the K-best elements}
   		$K^{\prime\prime}\gets\min\{K,M_c^{k}\}$\tcp*[r]{For $K>M_c^{k}$}   
   		$\mathcal{D}\gets \{\mathcal{D}_{\mathrm{sort}}\}_{1}^{K^{\prime\prime}}$ and $\mathcal{C}\gets \{\mathcal{C}_{\mathrm{sort}}\}_{1}^{K^{\prime\prime}}$\;
   		\tcp*[l]{Update the exponential term}
   		$\mathcal{E}\gets\mathcal{E} + \sum_{c\in{\mathcal{C}_{\mathrm{sort}}}\backslash{\mathcal{C}}}\exp(-c)\cdot M_c^{N_t-k}$\;   		
   		\tcp*[l]{Go to next depth}
   		\BFS{$\{\mathbf{v},\mathbf{R},K\},\{k+1,\mathcal{D},\mathcal{C},\mathcal{E}\}$}\;
   }
\KwRet{$\mathcal{D}_{\mathrm{BFS}}^{(K)}, \mathcal{E}$}
}
\end{algorithm}

\section{SNR-Based Algorithmic Extension}\label{sec:SD_approximation}

\begin{figure*}[tb]
  \centering
  \subfigure[]{\includegraphics[width=2.3in]{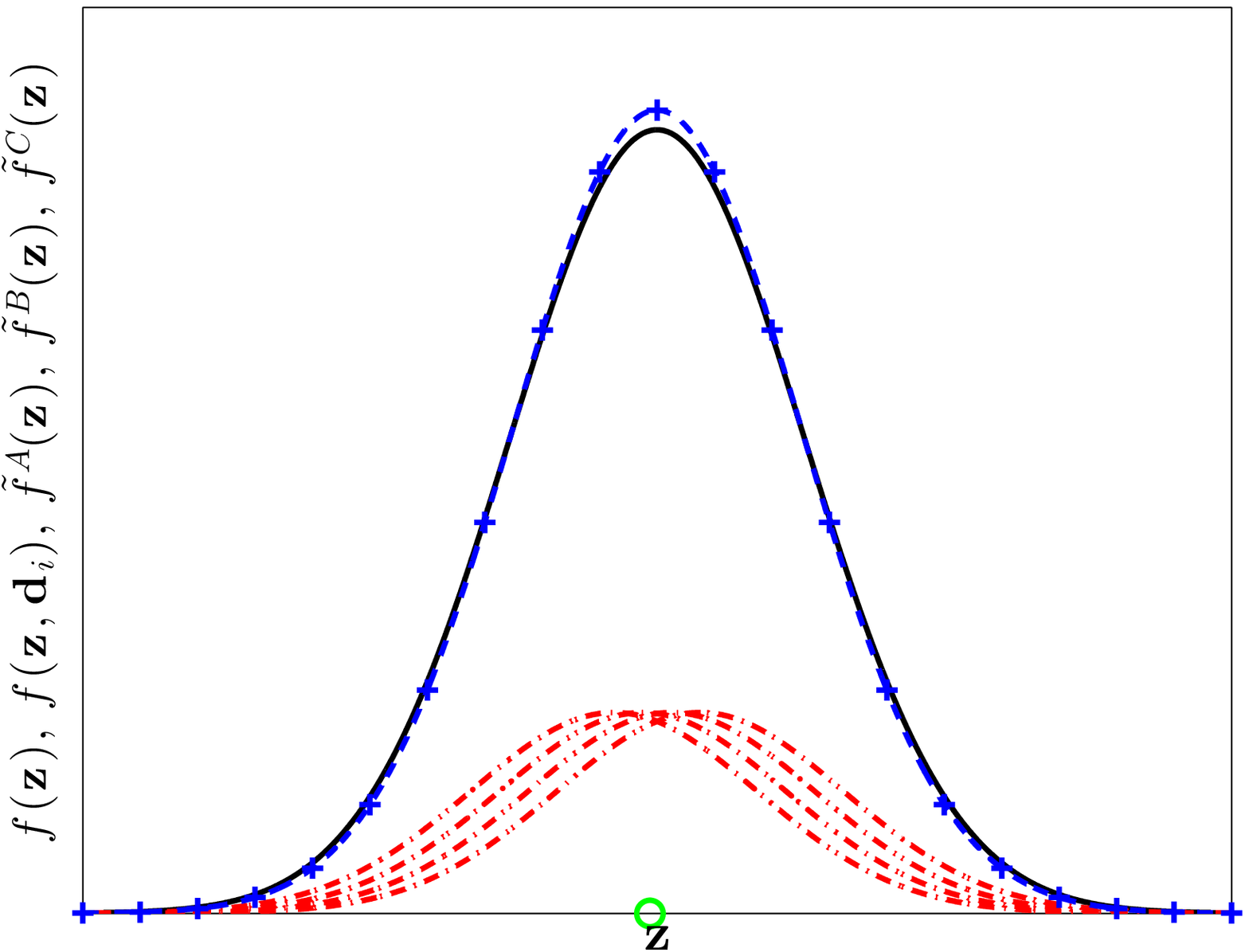}}
  \subfigure[]{\includegraphics[width=2.3in]{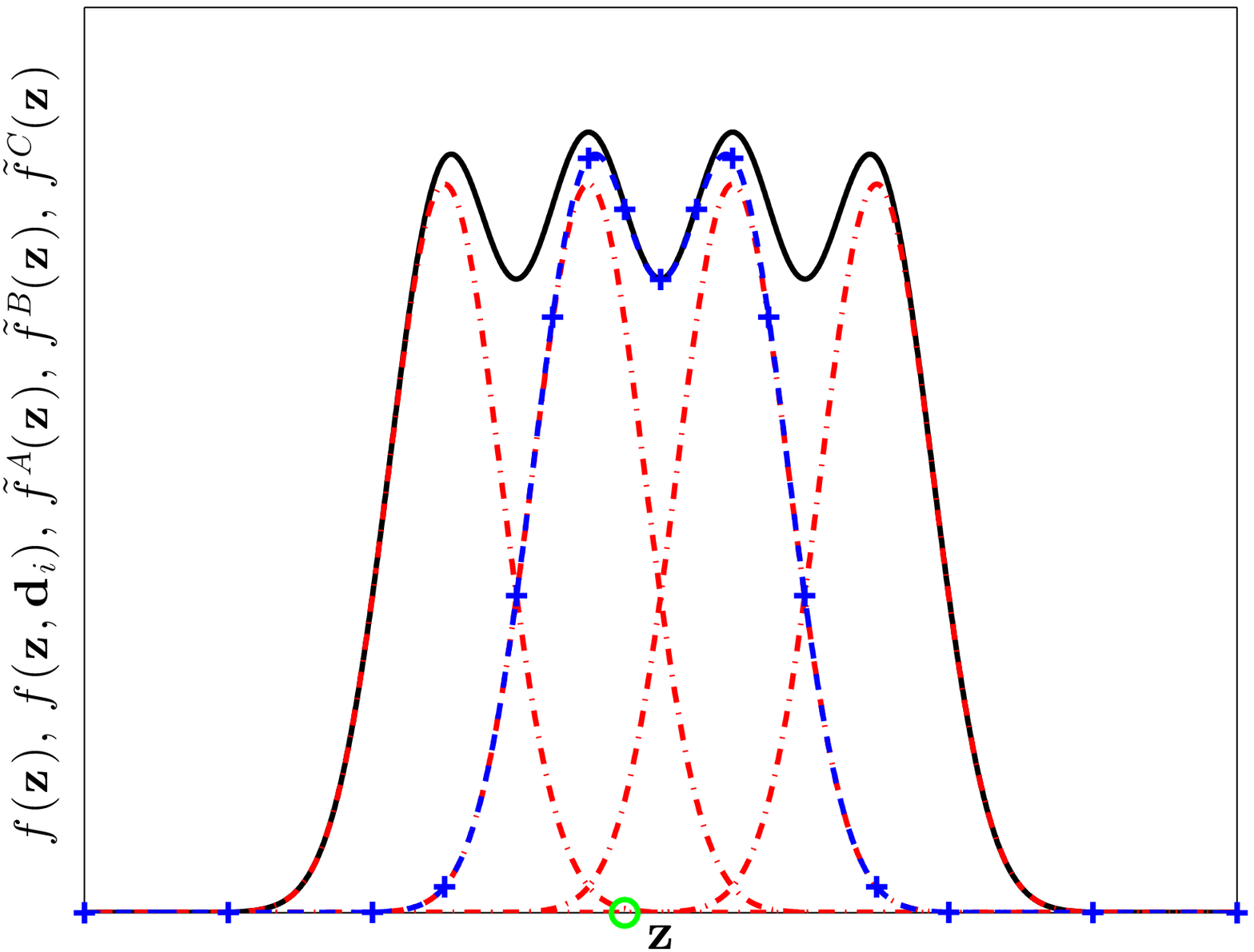}}
  \subfigure[]{\includegraphics[width=2.3in]{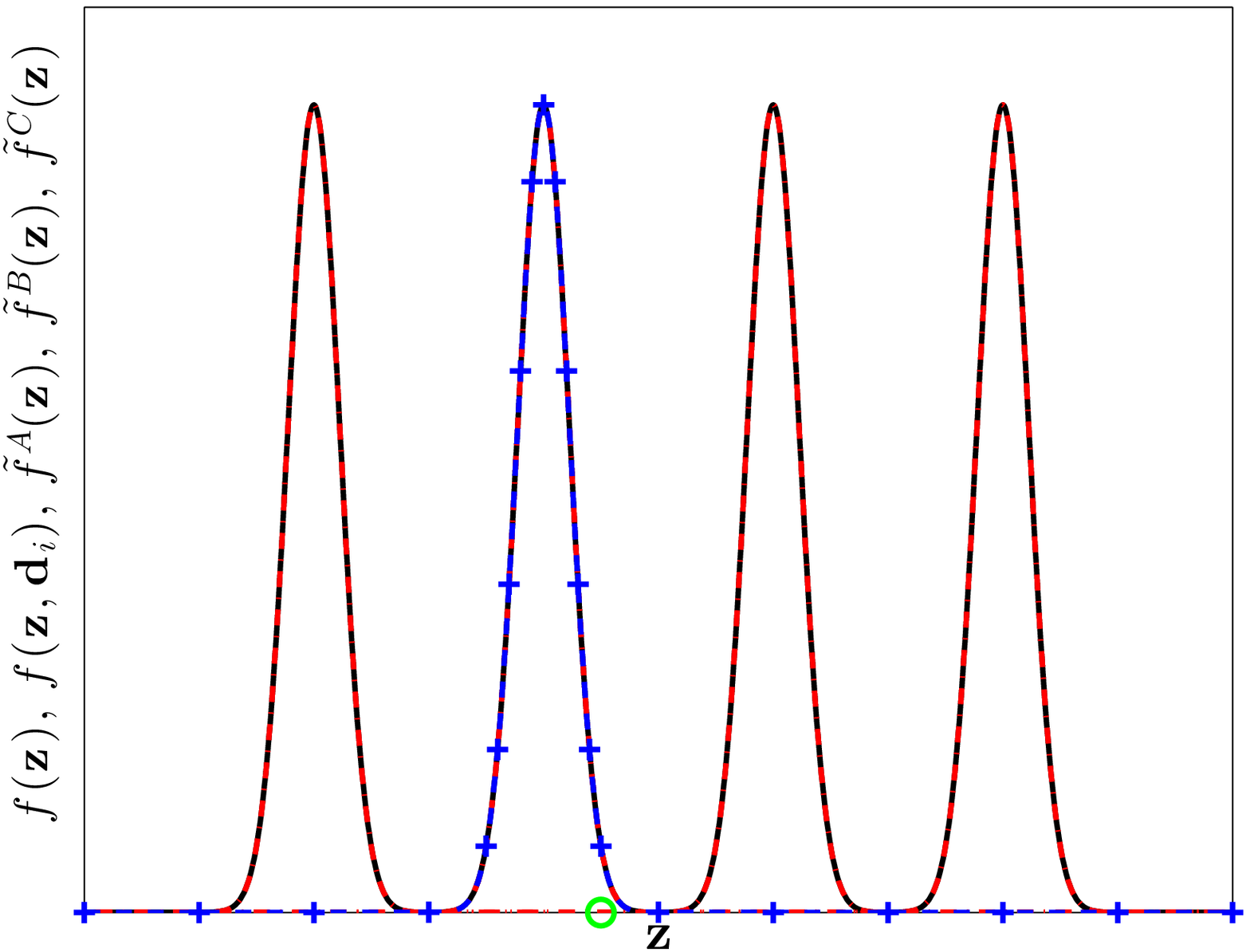}}
  \caption{An example of three approximations according to SNR region (e.g., 4-PAM): (a) low SNR -- single Gaussian approximation, $\tilde{f}^{A}(\mathbf{z})$. (b) medium SNR -- 2-closest components approximation based on the SD tree search, $\tilde{f}^{B}(\mathbf{z})$. (c) high SNR -- Babai estimate-based approximation, $\tilde{f}^{C}(\mathbf{z})$. The red dashed-dotted lines denote the pdfs of four different Gaussian components, $f(\mathbf{z},\mathbf{d}_i)=p(\mathbf{d}_i)f(\mathbf{z}|\mathbf{d}_i)$, the black line denotes the pdf of the true Gaussian mixture, $f(\mathbf{z})$, and the blue dashed line with `+' marker denotes the approximated pdf. The green circle denotes the drawn $\mathbf{z}$ in Monte Carlo method, for which $f(\mathbf{z})$ has to be approximated.}
\label{fig:three_SNR_region}
\end{figure*}

The complexity of the previous algorithms may be still too high for a large number of components. 
In the following subsection, we propose another approach to further reduce the complexity significantly.
For a given complexity, the approach can be also used to improve the precision by increasing the number of considered components in the range what it matters.

The main idea of the extension is to apply different approximation methods to partial symbol vectors within different SNR regions and combine them in order to compute the entropy in the mutual information. To this end, we first partition the given channel matrix and input symbol vector to three regions with respect to the SNR: \REV{(i) \emph{low SNR}, (ii) \emph{medium SNR}, and (iii) \emph{high SNR}.}
Thereafter, we apply \emph{one component only approximation}, the \emph{SD upper bound}, and \emph{single Gaussian approximation}, respectively. Finally, we combine them over the unified symbol vector.
Fig.~\ref{fig:three_SNR_region} illustrates a simple 4-pulse amplitude modulation (PAM) example of three different approximation methods suitable for different SNR. 
In the figure, each approximated pdf is well-matched with the true Gaussian mixture pdf with respect to the drawn $\mathbf{z}$ in Monte Carlo method. 
This is the main motivation of this SNR region based approximation in this section.

According to the above partitioning, the received signal model \eqref{eq:sys_eq_qr} can be rewritten as
\begin{align}
\left[ \begin{array}{c}
\mathbf{v}_A \\ 
\mathbf{v}_B \\ 
\mathbf{v}_C
\end{array}
\right] = 
\left[ \begin{array}{ccc}
\mathbf{A}_{} & \mathbf{B}_{A} & \mathbf{C}_{A} \\ 
0 & \mathbf{B}_{} & \mathbf{C}_{B} \\ 
0 & 0 & \mathbf{C}_{}
\end{array} \right] 
\left[ \begin{array}{c}
\mathbf{d}_A \\ 
\mathbf{d}_B \\ 
\mathbf{d}_C
\end{array}
\right]
+
\left[ \begin{array}{c}
\mathbf{w}_A \\ 
\mathbf{w}_B \\ 
\mathbf{w}_C
\end{array}
\right],
\label{eq:system_novel_approx}
\end{align}
where $\mathbf{A}\in\mathbb{C}^{N_A\times N_A}$, $\mathbf{B}\in\mathbb{C}^{N_B\times N_B}$, and $\mathbf{C}\in\mathbb{C}^{N_C\times N_C}$ in which $N_t = N_A + N_B + N_C$.
Let $\mathrm{diag}(\mathbf{R})=[\lambda_1, \ldots, \lambda_{N_t}]$.
Assuming the diagonal terms in $\mathbf{R}$ are ordered in increasing order, the following relations hold with respect to two threshold values, $\gamma_l$ and $\gamma_h$:
\begin{align}
\underbrace{\lambda_{1}^2 \leq \ldots \leq \lambda_{N_A}^2}_{\textrm{low SNR}} &\leq \gamma_l < \underbrace{\lambda_{N_A+1}^2 \leq \ldots \leq \lambda_{N_A + N_B}^2}_{\textrm{medium SNR}}\nonumber\\
&\leq \gamma_h < \underbrace{\lambda_{N_A+N_B+1}^2 \leq \ldots \leq \lambda_{N_t}^2}_{\textrm{high SNR}}. \label{eq:partition_eigenvalue}
\end{align}
Consequently, $\mathbf{A}$, $\mathbf{B}$, and $\mathbf{C}$ in \eqref{eq:system_novel_approx}  correspond to low SNR, medium SNR, and high SNR partitions, respectively, after reordering the original channel matrix, i.e., $\tilde{\mathbf{H}} = \mathbf{H\Pi}$ where $\mathbf{\Pi}$ is the permutation matrix, such that the eigenvalues are sorted in increasing order. 
The V-BLAST ZF-DFE channel ordering in \cite{DGC03TIT} provides an eigenvalue ordering method. 
Note that the sorting may not be perfect but it is sufficiently good for our purpose since the differences are small.
Similarly to $\alpha$ and $K$ parameters, $\gamma_l$ and $\gamma_h$ are design parameters which trade off accuracy versus complexity.
\REV{At medium SNR, both parameters need to be carefully chosen since they can still cause a prohibitive computational complexity.}
A discussion on the choice of those parameters is provided in Section~\ref{subsec:gamma_ths}.

\subsection{SNR-Based Enhanced Approximation}

In this subsection, we propose an SNR-based extension of the SD approximation method. Therefore, we first present three approximation methods for three difference SNR partition. Then, we provide the approximated pdf combining those results.

We start from the high SNR partition corresponding to the block $\mathbf{C}$.
The effective received signal at high SNR can be approximated by
\begin{align}
\mathbf{v}_C &= \mathbf{C} \mathbf{d}_C + \mathbf{w}_C
\approx \mathbf{C} \tilde{\mathbf{d}}_{C} + \mathbf{w}_C,
\label{eq:v_C}
\end{align}
where $\tilde{\mathbf{d}}_{C}$ is the drawn $\mathbf{d}_C$ in the Monte-Carlo method, thus it is known to us for the computation. At high SNR, this approximation becomes very good due to negligible noise as shown in Fig.~\ref{fig:three_SNR_region}~(c).

By applying the known component for the high SNR block, the effective received signal at medium SNR is approximated by
\begin{align}
\mathbf{v}_B &= \mathbf{B} \mathbf{d}_B + \mathbf{C}_B \mathbf{d}_{C}  + \mathbf{w}_B\nonumber\\
&\approx \mathbf{B} \mathbf{d}_B + \mathbf{C}_B \tilde{\mathbf{d}}_{C}  + \mathbf{w}_B.
\label{eq:v_B}
\end{align}
For given $\tilde{\mathbf{d}}_{C}$, we have
\begin{align}
\mathbf{v}_B^{\prime} &= \mathbf{v}_B - \mathbf{C}_B \tilde{\mathbf{d}}_{C}
\approx \mathbf{B} \mathbf{d}_B + \mathbf{w}_B.
\label{eq:v_B_p}
\end{align}
Similarly as in the previous sections, we apply either the DFS-based tree search or the BFS-based tree search to \eqref{eq:v_B_p} instead of \eqref{eq:sys_eq_qr}.
For the DFS-based tree search, the sphere radius is set to $\zeta^2 = \alpha \| \mathbf{v}_B^{\prime} - \mathbf{B} \mathbf{d}_{0,B} \|^2$ where $\mathbf{d}_{0,B}$ is the Babai estimate corresponding to $\mathbf{d}_{B}$.
For the BFS-based tree search, $\zeta^2$ is set to a sufficiently large value and the $K$ parameter is chosen considering the block size $N_B$.
Afterwards, we can find the vector set $\mathcal{D}_{B}^{\mathrm{SD}}=\{\hat{\mathbf{d}}_{B,1}, \hat{\mathbf{d}}_{B,2}, \ldots, \hat{\mathbf{d}}_{B,|\mathcal{D}_{B}^{\mathrm{SD}}|}\}$ where either $\mathcal{D}_{B}^{\mathrm{SD}}=\mathcal{D}_{\mathrm{DFS}}^{(\zeta)}$ if the DFS-based tree search is used or $\mathcal{D}_{B}^{\mathrm{SD}}=\mathcal{D}_{\mathrm{BFS}}^{(K)}$ if the BFS-based tree search is used.

Similarly to the medium SNR case, by applying the Babai estimate for the high SNR block, the effective received signal at low SNR is given by
\begin{align}
\mathbf{v}_A &= \mathbf{A} \mathbf{d}_A + \mathbf{B}_A \mathbf{d}_B + \mathbf{C}_A \mathbf{d}_C + \mathbf{w}_A \nonumber\\
&\approx \mathbf{A} \mathbf{d}_A + \mathbf{B}_A \mathbf{d}_{B}  + \mathbf{C}_A \tilde{\mathbf{d}}_{C} + \mathbf{w}_A.
\label{eq:v_A}
\end{align}
For given $\tilde{\mathbf{d}}_{C}$, we have
\begin{align}
\mathbf{v}_A^{\prime} &= \mathbf{v}_A - \mathbf{C}_A \tilde{\mathbf{d}}_{C} 
\approx \mathbf{A} \mathbf{d}_A + \mathbf{B}_A \mathbf{d}_B + \mathbf{w}_A. 
\label{eq:v_A_p}
\end{align}
For each of the $|\mathcal{D}_{B}^{\mathrm{SD}}|$-closest vectors, $\hat{\mathbf{d}}_{B}\in\mathcal{D}_{B}^{\mathrm{SD}}$,
we have
 \begin{align}
\mathbf{v}_{A}^{\prime} = \mathbf{A} \mathbf{d}_A + \mathbf{B}_A \hat{\mathbf{d}}_{B} + \mathbf{w}_A. 
\label{eq:v_A_p}
\end{align}
Hence, for given $\hat{\mathbf{d}}_{B}$, we arrive at
\begin{align}
\mathbf{v}_{A}^{\prime\prime} &= \mathbf{v}_{A}^{\prime} - \mathbf{B}_A\hat{\mathbf{d}}_{B} = \mathbf{A}\mathbf{d}_A  + \mathbf{w}_A,
\label{eq:v_A_pp}
\end{align}
which follows a Gaussian mixture distribution similar to \eqref{eq:v_B_p}.
For each given $\hat{\mathbf{d}}_{B,m}$, we approximate the Gaussian mixture distribution $f( \mathbf{v}_{A,m}^{\prime\prime})$ by a single Gaussian distribution with same mean and covariance for the low SNR block $\mathbf{A}$ as shown in Fig.~\ref{fig:three_SNR_region}~(a).

Applying the three different approximations to the three SNR partition, the pdf of the unified received symbol vector can be derived as
\begin{align}
f(\mathbf{v}) 
&= f(\mathbf{v}_C,\mathbf{v}_B,\mathbf{v}_A)= f(\mathbf{v}_C)f(\mathbf{v}_B,\mathbf{v}_A|\mathbf{v}_C)\nonumber\\
&= \sum_{\hat{\mathbf{d}}_C\in\mathcal{D}_C}p(\hat{\mathbf{d}}_C)f(\mathbf{v}_C|\hat{\mathbf{d}}_C)f(\mathbf{v}_B,\mathbf{v}_A|\mathbf{v}_C,\hat{\mathbf{d}}_C)\nonumber\\
&\overset{(\mathrm{a})}{\geq} p(\tilde{\mathbf{d}}_{C})f(\mathbf{v}_C|\tilde{\mathbf{d}}_{C})f(\mathbf{v}_B,\mathbf{v}_A|\mathbf{v}_C,\tilde{\mathbf{d}}_{C})\nonumber\\
&\overset{(\mathrm{b})}{\approx} p(\tilde{\mathbf{d}}_{C})f(\mathbf{v}_C|\tilde{\mathbf{d}}_{C})f(\mathbf{v}_B,\mathbf{v}_A|\tilde{\mathbf{d}}_{C})\nonumber\\
&= p(\tilde{\mathbf{d}}_{C})f(\mathbf{v}_C|\tilde{\mathbf{d}}_{C})f(\mathbf{v}_B|\tilde{\mathbf{d}}_{C})f(\mathbf{v}_A|\mathbf{v}_B,\tilde{\mathbf{d}}_{C})\nonumber\\
&= p(\tilde{\mathbf{d}}_{C})f(\mathbf{v}_C|\tilde{\mathbf{d}}_{C})\cdot\nonumber\\
&\quad\Big[\sum_{\hat{\mathbf{d}}_B\in\mathcal{D}_B}p(\hat{\mathbf{d}}_B)f(\mathbf{v}_B|\tilde{\mathbf{d}}_{C},\hat{\mathbf{d}}_B)f(\mathbf{v}_A|\mathbf{v}_B,\tilde{\mathbf{d}}_{C},\hat{\mathbf{d}}_B)\Big]\nonumber\\
&\overset{(\mathrm{c})}{\geq} p(\tilde{\mathbf{d}}_{C})f(\mathbf{v}_C|\tilde{\mathbf{d}}_{C})\cdot\nonumber\\
&\quad\Big[\sum_{\hat{\mathbf{d}}_B\in\mathcal{D}_B^{\mathrm{SD}}}p(\hat{\mathbf{d}}_B)f(\mathbf{v}_B|\tilde{\mathbf{d}}_{C},\hat{\mathbf{d}}_B)f(\mathbf{v}_A|\mathbf{v}_B,\tilde{\mathbf{d}}_{C},\hat{\mathbf{d}}_B)\Big]\nonumber\\
&\overset{(\mathrm{d})}{\approx} p(\tilde{\mathbf{d}}_{C})f(\mathbf{v}_C|\tilde{\mathbf{d}}_{C})\cdot\nonumber\\
&\quad\Big[\sum_{\hat{\mathbf{d}}_B\in\mathcal{D}_B^{\mathrm{SD}}}p(\hat{\mathbf{d}}_B)f(\mathbf{v}_B|\tilde{\mathbf{d}}_{C},\hat{\mathbf{d}}_B)f(\mathbf{v}_A|\tilde{\mathbf{d}}_{C},\hat{\mathbf{d}}_B)\Big]\nonumber\\
&= p(\tilde{\mathbf{d}}_{C})f(\mathbf{v}_C|\tilde{\mathbf{d}}_{C})\Big[\sum_{\hat{\mathbf{d}}_B\in\mathcal{D}_B^{\mathrm{SD}}}p(\hat{\mathbf{d}}_B)f(\mathbf{v}_B|\tilde{\mathbf{d}}_{C},\hat{\mathbf{d}}_B)\cdot\nonumber\\
&\quad\sum_{\hat{\mathbf{d}}_A\in\mathcal{D}_A}p(\hat{\mathbf{d}}_A)f(\mathbf{v}_A|\tilde{\mathbf{d}}_{C},\hat{\mathbf{d}}_B,\hat{\mathbf{d}}_A)\Big]\nonumber\\
&\overset{(\mathrm{e})}{\approx} p(\tilde{\mathbf{d}}_{C})f(\mathbf{v}_C|\tilde{\mathbf{d}}_{C})\cdot\nonumber\\
&\quad\Big[\sum_{\hat{\mathbf{d}}_B\in\mathcal{D}_B^{\mathrm{SD}}}p(\hat{\mathbf{d}}_B)f(\mathbf{v}_B|\tilde{\mathbf{d}}_{C},\hat{\mathbf{d}}_B)f_G(\mathbf{v}_A|\tilde{\mathbf{d}}_{C},\hat{\mathbf{d}}_B)\Big],\label{eq:f_v}
\end{align}
where (a) is the single component-based approximation, (c) is the SD upper bound, (e) is the single Gaussian approximation, and (b) and (d) follow from \eqref{eq:v_B} and \eqref{eq:v_A}.
In \eqref{eq:f_v}, each term is given by $p(\tilde{\mathbf{d}}_{C})=\frac{1}{M_c^{N_C}}$, $p(\hat{\mathbf{d}}_B)=\frac{1}{M_c^{N_B}}$,
\begin{align}
&f(\mathbf{v}_C|\tilde{\mathbf{d}}_{C})=\frac{1}{\pi^{N_C}}\exp\left(-\|\mathbf{v}_C - \boldsymbol{\mu}_C\|^2\right),\label{eq:f_v_C_d_0C}\\
&f(\mathbf{v}_B|\tilde{\mathbf{d}}_{C},\hat{\mathbf{d}}_B)=\frac{1}{\pi^{N_B}}\exp\left(-\|\mathbf{v}_B - \boldsymbol{\mu}_B\|^2\right),\label{eq:f_v_B_d_0C_hd_B}\\
&f_G(\mathbf{v}_A|\tilde{\mathbf{d}}_{C},\hat{\mathbf{d}}_B)\nonumber\\
&=\frac{1}{\pi^{N_A}\det{\mathbf{K}_A}}\exp\left(-(\mathbf{v}_A - \boldsymbol{\mu}_A)^{\mathsf{H}}\mathbf{K}_A^{-1}(\mathbf{v}_A - \boldsymbol{\mu}_A)\right),\label{eq:f_v_A_d_0C_hd_B}
\end{align}
where $\boldsymbol{\mu}_C = \mathbf{C}\tilde{\mathbf{d}}_{C}$, $\boldsymbol{\mu}_B = \mathbf{B}\hat{\mathbf{d}}_B + \mathbf{C}_B\tilde{\mathbf{d}}_{C}$, $\boldsymbol{\mu}_A = \mathbf{B}_A\hat{\mathbf{d}}_B + \mathbf{C}_A\tilde{\mathbf{d}}_{C}$, and $\mathbf{K}_A = \rho\mathbf{AA}^{\mathsf{H}}+\mathbf{I}$.
Note that this novel approximation can reduce the tree search complexity from $M_c^{N_t}$ to $M_c^{N_B}$ where $N_B$ is determined by both $\gamma_l$ and $\gamma_h$ parameters.
Since (a) and (c) give lower bounds, when (b), (d), and (e) are very accurate approximations, the final pdf in \eqref{eq:f_v} can be a lower bound (equivalently, an upper bound on the entropy). However, in general, it is an approximation due to (b), (d), and (e).
The overall SNR-based enhanced approximation algorithm is presented in Algorithm~\ref{alg:SD_app}.\\

\begin{algorithm}[t]\label{alg:SD_app}
  \DontPrintSemicolon
   \caption{SNR-Based Enhanced Approximation}
   \KwIn{$\mathbf{H},\rho$}
   \KwOut{$\hat{h}_{\mathrm{SDEA}}$}   
   \BlankLine
   \textbf{Initialization:} Set $\gamma_l$ and $\gamma_h$\;
   ${\tilde{\mathbf{H}}}={\mathbf{H}}\mathbf{\Pi}$ according to \cite{DGC03TIT}\tcp*[l]{Channel ordering}
   $[\mathbf{Q}~\mathbf{R}]\gets \mathrm{qr}(\tilde{\mathbf{H}})$\tcp*[r]{QR factorization}
   \tcp*[l]{Channel matrix partition}
   Find $\mathbf{A}$, $\mathbf{B}$, $\mathbf{C}$, $\mathbf{B}_{A}$, $\mathbf{C}_{A}$, and $\mathbf{C}_{B}$ according to \eqref{eq:partition_eigenvalue}\;
   \tcp*[l]{Integration by a Monte-Carlo method}
   \For(\tcp*[f]{Loop for $\mathbf{d}$}){$i = 1$ \KwTo $N_{d}$}
   {
      Generate $\mathbf{d}^{(i)} \gets \sqrt{\rho}\cdot \mathbf{s}$ where $\mathbf{s}\sim\mathcal{U}(\mathcal{M}^{N_t})$\;
      \For(\tcp*[f]{Loop for $\mathbf{n}$}){$j = 1$ \KwTo $N_{n}$}
      {
      	Generate $\mathbf{n}^{(j)}$ where $\mathbf{n}^{(j)}\sim\mathcal{CN}(0,\mathbf{I})$\;
		$\mathbf{z}^{(i,j)} \gets \tilde{\mathbf{H}}\mathbf{d}^{(i)} + \mathbf{n}^{(j)}$\;
		$\mathbf{v}^{(i,j)} \gets \mathbf{Q}^{\mathsf{H}}\mathbf{z}^{(i,j)}$\;
		Find $\mathbf{d}_{A}^{(i)}$, $\mathbf{d}_{B}^{(i)}$, $\mathbf{d}_{C}^{(i)}$, $\mathbf{v}_{A}^{(i,j)}$, $\mathbf{v}_{B}^{(i,j)}$, and $\mathbf{v}_{C}^{(i,j)}$\;
		\tcp*[l]{Babai estimate}
		$\mathbf{d}_{0} \gets \tilde{\mathbf{H}}^{\dagger}\mathbf{z}^{(i,j)}$ and find $\mathbf{d}_{0,A}$, $\mathbf{d}_{0,B}$, and $\mathbf{d}_{0,C}$\;
		\tcp*[l]{(C) High SNR approximation}
		Compute $f(\mathbf{v}_C^{(i,j)}|\mathbf{d}_{0,C})$ according to \eqref{eq:f_v_C_d_0C}\;  
		\tcp*[l]{(B) Medium SNR approximation}
		$\mathbf{v}_B^{\prime (i,j)} \gets \mathbf{v}_B^{(i,j)} - \mathbf{C}_B \mathbf{d}_{C}^{(i)}$\;  		
		\tcp*[l]{Call SD tree search algorithm}
		\uIf{$\mathrm{DFS}$}{
			Set $\alpha\geq 1$ and		
			$\zeta^{2} \gets \alpha \| \mathbf{v}_B^{\prime(i,j)} -  \mathbf{R}{\mathbf{d}}_{0,B}^{(j)} \|^2$\;  		
			$[\mathcal{D}_B^{\mathrm{SD}}, \mathcal{E}]\gets$~\DFS{$\{\mathbf{v}_B^{\prime(i,j)},\mathbf{R},\zeta^2\}$,$\{1,[~],0,0,\emptyset\}$}\
		}  
		\ElseIf{$\mathrm{BFS}$}{
			Set $K$ according to \eqref{eq:K_C0}\;  				
			$[\mathcal{D}_B^{\mathrm{SD}}, \mathcal{E}]\gets$~\BFS{$\{\mathbf{v}_B^{\prime(i,j)},\mathbf{R},K\}$,$\{1,\emptyset,\emptyset,0\}$}\;
		}  
		Compute $f(\mathbf{v}_{B}^{(i,j)}|\mathbf{d}_{0,C},\hat{\mathbf{d}}_B)$, $\forall \hat{\mathbf{d}}_B\in\mathcal{D}_B^{\mathrm{SD}}$, according to \eqref{eq:f_v_B_d_0C_hd_B}\;
		\tcp*[l]{(A) Low SNR approximation}      
		Compute $f(\mathbf{v}_{A}^{(i,j)}|\mathbf{d}_{0,C},\hat{\mathbf{d}}_{B})$, $\forall \hat{\mathbf{d}}_B\in\mathcal{D}_B^{\mathrm{SD}}$, according to \eqref{eq:f_v_A_d_0C_hd_B}\;
		\tcp*[l]{Compute the pdf of $\mathbf{v}$}
		Compute $f(\mathbf{v}^{(i,j)})$ according to \eqref{eq:f_v}\; 
      }
      \tcp*[l]{Compute entropy approximation}
      $\hat{h}_{\mathrm{SDEA}}=-\frac{1}{N_d N_n}\sum_{i=1}^{N_d}\sum_{j=1}^{N_n}\log_2 f(\mathbf{v}^{(i,j)}) $\;
   }
\end{algorithm}

\subsection{Discussion on $\gamma_l$ and $\gamma_h$ Parameters}\label{subsec:gamma_ths}
Since $\gamma_l$ and $\gamma_h$ parameters determine the size of the submatrix $\mathbf{B}$, they highly influence the complexity reduction gain.
Basically, if the difference between those parameters is small, the proposed approximation yields low complexity with some accuracy losses. On the contrary, as the difference increases, it converges to the SD upper bound results. The goal is to set the parameters so that the accuracy losses are still acceptable. In this subsection, we investigate trends of accuracy on the mutual information according to those parameters which will provide us with \REV{a guideline how to determine them.}

\begin{figure}[tb]
  \centering
  \includegraphics[width=6.5cm]{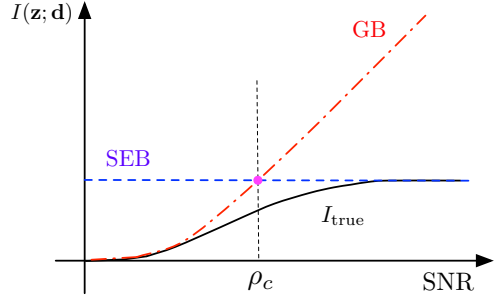}
  \caption{The relation among the GB, the SEB, and the true mutual information according to SNR. $\rho_c$ denotes the SNR corresponding to the intersection of the GB and the SEB.}
\label{fig:GB_SEB_True}
\vspace{-0.4cm}
\end{figure}

Even though we focus on the entropy approximation, our main results are evaluated in terms of the mutual information in Section~\ref{sec:numerical_examples}.
Thus, we determine the parameters based on the mutual information in this subsection.
First of all, there exist two trivial upper bounds on the mutual information: (i) \emph{Gaussian bound} (GB) assuming Gaussian input distribution given by $I(\mathbf{z};\mathbf{d}) = \log_2\det\left(\mathbf{I}+\rho\mathbf{H}\mathbf{H}^{\mathsf{H}}\right)$; (ii) \emph{source entropy bound} (SEB) such that the mutual information cannot exceed the source entropy, i.e., $I(\mathbf{z};\mathbf{d}) = H(\mathbf{d}) - H(\mathbf{d}|\mathbf{z}) \leq H(\mathbf{d})=\log_2 M_c^{N_t}$ since the entropy is non-negative. Fig.~\ref{fig:GB_SEB_True} shows a typical relation among the GB, the SEB, and the true mutual information according to SNR.

\begin{figure}[tb]
  \centering
  \subfigure[]{\includegraphics[width=8cm]{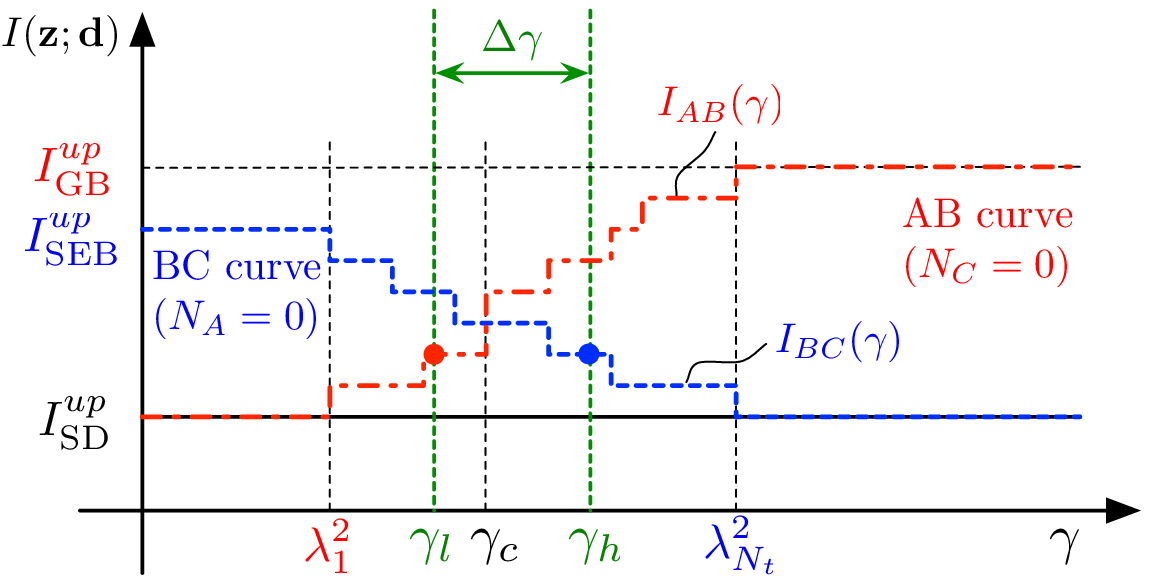}}
  \subfigure[]{\includegraphics[width=8cm]{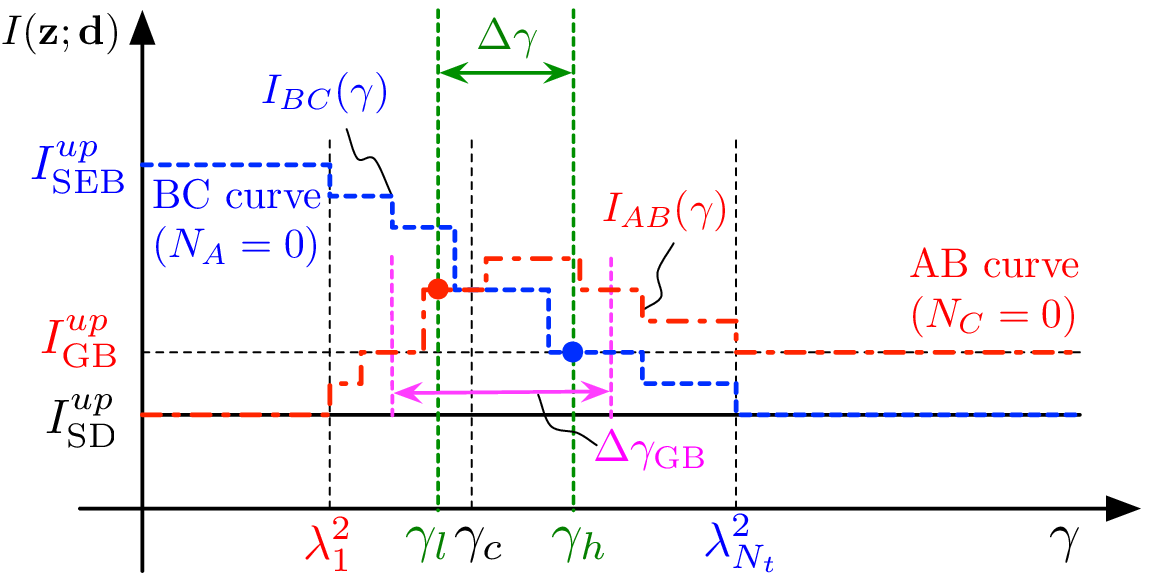}}
  \caption{Mutual information according to $\gamma_l$ and $\gamma_h$ at given SNR $\rho$: (a) $\rho > \rho_c$ (i.e., $I_{\mathrm{SEB}}^{up} < I_{\mathrm{GB}}^{up}$) case (b) $\rho < \rho_c$ (i.e., $I_{\mathrm{SEB}}^{up} >  I_{\mathrm{GB}}^{up}$) case. \REV{The thick blue dashed curve is for fixing $\gamma_l \rightarrow 0$ and varying $\gamma_h$ (`BC curve') and the thick red dot-and-dash curve is for fixing $\gamma_h \rightarrow \infty$ and varying $\gamma_l$ (`AB curve').} $I_{\mathrm{AB}}(\gamma_l)-I_{\mathrm{SD}}^{up}$ and $I_{\mathrm{BC}}(\gamma_h)-I_{\mathrm{SD}}^{up}$ correspond to approximation errors.}
\label{fig:gamma_threshold}
\vspace{-0.4cm}
\end{figure}

Let $I_{\mathrm{GB}}^{up}$, $I_{\mathrm{SEB}}^{up}$, $I_{\mathrm{SD}}^{up}$, and $I_{\mathrm{SDEA}}$ denote GB, SEB, SD upper bound, and SD-based enhanced approximation on the mutual information, respectively.
Through numerical observations, the basic trends of the mutual information according to $\gamma_l$ and $\gamma_h$ parameters for given SNR are illustrated in Fig.~\ref{fig:gamma_threshold}.
In the figures, we draw two mutual information curves fixing one threshold and varying the other: \REV{The thick blue dashed curve is for fixing $\gamma_l \rightarrow 0$ (equivalently, $\gamma_l < \lambda_{1}^2$) and varying $\gamma_h$ ($N_A=0$ case, so called `BC curve'); The thick red dot-and-dash curve is for fixing $\gamma_h \rightarrow \infty$ (equivalently, $\gamma_h > \lambda_{N_t}^2$) and varying $\gamma_l$ ($N_C=0$ case, so called `AB curve').}

The properties of the mutual information of the SNR-based enhanced approximation, $I_{\mathrm{SDEA}}$, on $\gamma_l$ and $\gamma_h$ are as follows:
\begin{itemize}
\item[a)] If $\gamma_l \leq \gamma_h < \lambda_{1}^2$, $I_{\mathrm{SDEA}}= I_{\mathrm{SEB}}^{up}$.
\item[b)] If $\gamma_h > \gamma_l \geq \lambda_{N_t}^2$, $I_{\mathrm{SDEA}}= I_{\mathrm{GB}}^{up}$.
\item[c)] If $\gamma_l < \lambda_{1}^2$ and $\gamma_h \geq \lambda_{N_t}^2$, $I_{\mathrm{SDEA}}= I_{\mathrm{SD}}^{up}$.
\item[d)] The BC curve monotonically decreases from $I_{\mathrm{SEB}}^{up}$ to $I_{\mathrm{SD}}^{up}$ as $\gamma_h$ increases from $\lambda_1^2$ to $\lambda_{N_t}^2$.
\item[e)] The AB curve can exceed $I_{\mathrm{GB}}^{up}$ at low SNR.
\end{itemize}
The proofs of the properties are provided in Appendix~\ref{app:Proof_of_Properties}.

Based on these properties, we next present a proposal how to determine $\gamma_l$ and $\gamma_h$ at given average SNR. 
Let us define $\Delta\gamma\triangleq \gamma_h - \gamma_l$. 
If $\Delta\gamma > \lambda_{N_t}^2 - \lambda_{1}^2$, it results in $I_{\mathrm{SD}}^{up}$ by setting such that Property~c can be satisfied. Otherwise, we trade off accuracy versus complexity.
As shown in Fig.~\ref{fig:gamma_threshold}, there exists an intersection point of the BC curve and the AB curve.
Let us denote the intersection point on $x$-axis by $\gamma_c$, then the AB curve is less erroneous on its left-hand side and so is the BC curve on its right-hand side of $\gamma_c$.
Hence, the best way is following the AB curve in the left-hand side and the BC curve in the right-hand side.
We propose to determine $\gamma_l$ and $\gamma_h$ proportionally to $\gamma_c - \lambda_{1}^2$ and $\lambda_{N_t}^2-\gamma_c$ with the width $\Delta\gamma$, i.e., $\gamma_l = \gamma_c + \frac{\lambda_{1}^2-\gamma_c}{\lambda_{N_t}^2 - \lambda_{1}^2}\Delta\gamma$ and $\gamma_h = \gamma_c + \frac{\lambda_{N_t}^2-\gamma_c}{\lambda_{N_t}^2 - \lambda_{1}^2}\Delta\gamma$.
If it is hard to find $\gamma_c$ due to computational complexity for high dimension, it can be determined by $\gamma_c=\frac{\lambda_1^2 + \lambda_{N_t}^2}{2}$.

According to SNR region, there are two different cases with comparison between $I_{\mathrm{GB}}^{up}$ and $I_{\mathrm{SEB}}^{up}$ as shown in Fig.~\ref{fig:gamma_threshold}~(a) and (b). In case of $I_{\mathrm{SEB}}^{up} >  I_{\mathrm{GB}}^{up}$, the error can make the mutual information exceed $I_{\mathrm{GB}}^{up}$ if the width obtained by $I_{\mathrm{GB}}^{up}$ and two curves (i.e., $\Delta\gamma_{\mathrm{GB}}$) is longer than $\Delta\gamma$. 
In this case, taking the GB is better than the SD approximation for given $\Delta\gamma$. Thus, two threshold values are set to $\gamma_l=\lambda_{N_t}^2\leq\gamma_h$ for this case based on Property~b. Actually, since the GB is very close to the true curve at low SNR, this setting is reasonable.
This also can be done simply by limiting the mutual information of the enhanced approximation by $I_{\mathrm{GB}}^{up}$. 

In general, we can identify three SNR regions. In the low SNR regime, the Gaussian approximation performs well. In the high SNR regime, the “one component only” approximation performs well. Both do not perform well in the medium SNR regime where the more complex SD approximation yields good results. The previous discussion applies to a given average SNR. If we want to compute the entropy/mutual information for an average SNR range as depicted in Fig. 4, then in principle we have to compute the threshold values for every average SNR value. However, to reduce the complexity even further, we propose to compute the threshold values $\gamma_l$ and $\gamma_h$ for the average SNR $\rho_c$ where the source entropy bound and Gaussian bound intersect and then scale the thresholds for each average SNR value $\rho$ by $\frac{\rho_c}{ \rho}$, i.e., $\gamma_l\frac{\rho_c}{\rho}$ and  $\gamma_h\frac{\rho_c}{\rho}$ are used as threshold values.

\section{Numerical Examples}\label{sec:numerical_examples}

In this section, we evaluate the proposed SD bounds and SNR-based enhanced approximations in terms of the mutual information and the complexity, compared to several benchmarks, which are briefly introduced in the following subsection. We consider two kinds of channels for the performance comparisons: (i) \emph{finite impulse response (FIR) filter channel} and (ii) \emph{frequency-selective and time-selective fading channel}.

\subsection{Benchmarks}

\subsubsection{Statistical Approximation (SA) Method \cite{ZSF+03WOC}}

The SA method is analogous to a combination of high and low SNR approximations in the proposed SNR-based enhanced approximation.
That is, it finds the following two pdfs of the received symbol vector for high and low SNRs, respectively,
\begin{align}
f_h(\mathbf{z}) &= \frac{1}{M_c^{N_t}} \exp\left(-\|\mathbf{z} - \mathbf{H}\tilde{\mathbf{d}} \|^2\right),\\
f_l(\mathbf{z}) &= \frac{1}{\pi^{N_t}\det(\mathbf{K_z})} \exp\left(-\mathbf{z}^{\mathsf{H}}\mathbf{K_z}^{-1}\mathbf{z}\right),
\end{align}
where $\tilde{\mathbf{d}}$ denotes the drawn $\mathbf{d}$ in the Monte-Carlo expectation and $\mathbf{K_z}=\rho\mathbf{H}\mathbf{H}^{\mathsf{H}}+\mathbf{I}$.
Then, the pdf of $\mathbf{z}$ is approximated by
\begin{align}
f(\mathbf{z}) \approx \max\left\{f_h(\mathbf{z}),f_l(\mathbf{z})\right\}.
\end{align}

\subsubsection{BCJR Algorithm Based Computation Method \cite{ALV+06TIT}}

The BCJR algorithm based computation method has been invented to compute information rates for finite-state channels.
In this method, for given finite-state channel, the mutual information between \emph{very long} input and output sequences are defined as 
\begin{align}
I(\mathbf{z};\mathbf{d}) \triangleq -\frac{1}{n}\log_2 p(z^n) - h(\mathbf{z}|\mathbf{d}),
\end{align}
where $n$ is the sequence length and $z^n=(z_1, z_2, \ldots, z_n)$ denotes the output sequence.
Then, it finds $p(z^n)$ based on the forward sum-product algorithm \cite{KFL01TIT}.
By employing a state sequence $s_{0}^{n}=(s_0,s_1,\ldots,s_n)$ and denoting the input sequence $d^n=(d_1,d_2,\ldots,d_n)$, $p(z^n)$ can be computed by
\begin{align}
p(z^n) = \sum_{d^n}\sum_{s_{0}^{n}}p(d^n,z^n,s_0^n).
\label{eq:p_z_n}
\end{align}
Defining the state metric $\mu_k(s_k)\triangleq p(s_k,z^k)$ for the $k$-th symbol, the computation of \eqref{eq:p_z_n} is possible by computing the state metrics recursively as
\begin{align}
\mu_k(s_k) &= \sum_{d_k}\sum_{s_{k-1}} \mu_{k-1}(s_{k-1})p(d_k,z_k,s_k|s_{k-1}) \label{eq:mu_k_s_k}\\
&= \sum_{d^k}\sum_{s_0^{k-1}} p(d^k,z^k,s_0^k),
\end{align}
for $k=1,2,\ldots,n$. After all, \eqref{eq:p_z_n} is obtained by $p(z^n) = \sum_{s_n}\mu_n(s_n)$.

In order to reduce the computational complexity for channels with a large number of states, \eqref{eq:mu_k_s_k} can be modified to yield a lower bound on $p(z^n)$ by taking a subset of states at each $k$ stage. 
Let $\mathcal{S}_{k}^{\prime}$ be a subset of states at the $k$-th stage with $Q \triangleq |\mathcal{S}_{k}^{\prime}|$.
The recursion \eqref{eq:mu_k_s_k} can be modified to
\begin{align}
\mu_k(s_k) &= \sum_{d_k}\sum_{s_{k-1}\in\mathcal{S}_{k-1}^{\prime}} \mu_{k-1}(s_{k-1})p(d_k,z_k,s_k|s_{k-1}). \label{eq:mu_k_s_k_RS}
\end{align}
This yields an upper bound on $h(\mathbf{z})$ and thus it is called \emph{reduced-state upper bound (RSUB)} in \cite{ALV+06TIT}.
It is worth noting that reducing the number of states is a similar approach to reducing the number of candidate input vectors in the proposed SD approximation.

\subsubsection{Hamming Distance 1 (HD1) Based Approximation Method}

For the sake of performance comparison, we propose an HD1-based approximation method which is a simple Gaussian mixture reduction including the symbol vectors with Hamming distance one from a pre-chosen symbol vector. Here, we use the Babai estimate for the pre-chosen symbol vector. 
Hence, based on the Babai estimate $\mathbf{d}_0=[d_{0,1},\ldots,d_{0,N_t}]^{\mathsf{T}}$, the candidate symbol vectors are obtained by
\begin{align}
\hat{\mathbf{d}}_i^{(j)} = [d_{0,1}, \ldots, d_{0,i-1}, d_{i}^{(j)}, d_{0,i+1} ,\ldots ,d_{0,N_t}]^{\mathsf{T}},
\end{align}
where $d_{i}^{(j)} \in \mathcal{M}_c \backslash \{d_{0,i}\}$, $i=1,\ldots,N_t$, $j=1,\ldots,|\mathcal{M}_c|-1$.
Consequently, we obtain the set of symbol vectors to be added up by $\mathcal{D}_{\mathrm{HD1}}=\{\mathbf{d}_0\}\bigcup\{\hat{\mathbf{d}}_{i}^{(j)}\}_{i,j}$ and thus, the following pdf is obtained:
\begin{align}
f_m(\mathbf{z}) = \sum_{\hat{\mathbf{d}}\in\mathcal{D}_{\mathrm{HD1}}} \frac{1}{M_c^{N_t}} \exp\left(-\|\mathbf{z} - \mathbf{H}\hat{\mathbf{d}}\|^2\right).
\end{align}
Since $f_m(\mathbf{z})$ is good only at medium SNR, by combining high and low SNR approximations in the SA method, the pdf of $\mathbf{z}$ can be approximated by
\begin{align}
f(\mathbf{z}) \approx \max\left\{f_h(\mathbf{z}),f_m(\mathbf{z}),f_l(\mathbf{z})\right\}.
\end{align}

\subsection{FIR Filter Channel}

As first example, we consider a memory-10 FIR filter channel with i.u.d. binary input used in \cite{ALV+06TIT} as the largest memory case, i.e., $z_k = \sum_{l=0}^{10} g_l d_{k-l} + n_k$, where $g_{l} = \frac{1}{1+(i-5)^2}$.
For convenience in SNR calculation, the sum of squared channel coefficients is normalized by one.
In matrix representation, this channel can be constructed by a Toeplitz matrix with $N_t=11$ where each row has the same elements but circularly shifted (i.e., circulant matrix).
Unlike real-valued noise was considered in \cite{ALV+06TIT}, we consider complex-valued noise. 

\begin{figure}
    \centering
    \subfigure[]{
        \includegraphics[width=8.5cm]{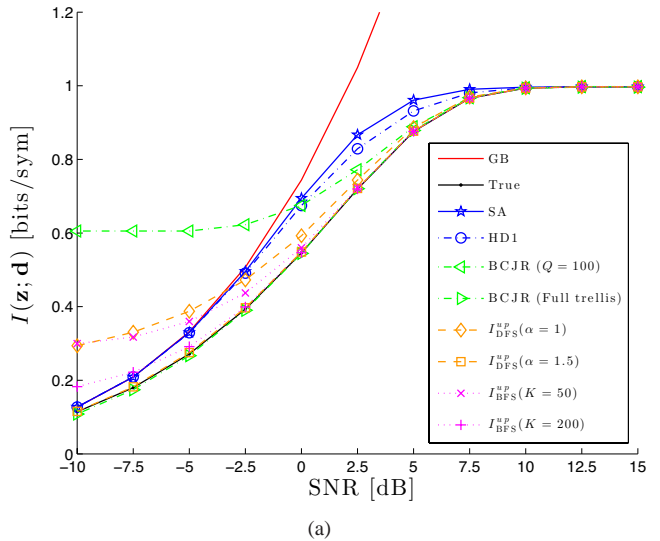}}
    \subfigure[]{
        \includegraphics[width=8.5cm]{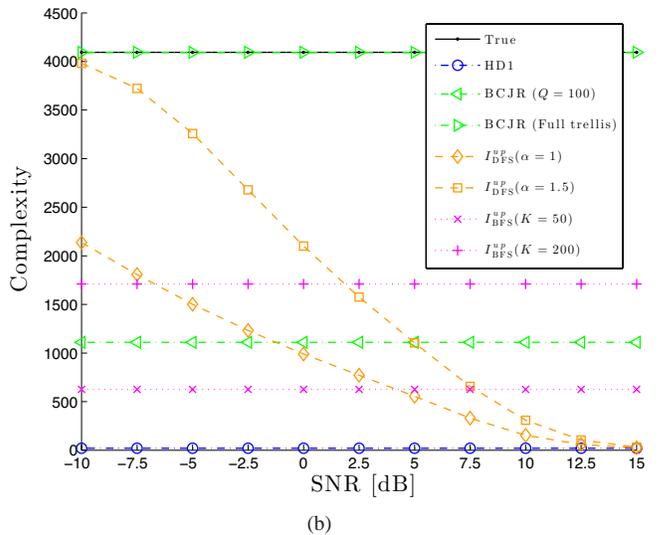}}
\caption{Memory-10 FIR filter channel with binary input (a) Mutual information [bits/symbol] (b) Complexity in terms of the number of visited nodes during the tree search or states during the trellis search. For Monte-Carlo expectation, we use $N_d=100$ and $N_z=50$, and for the BCJR method, we set $n=5\times10^4$.}
\label{fig:CH10_all}
\vspace{-0.4cm}
\end{figure}

Fig.~\ref{fig:CH10_all}~(a) shows the mutual information for memory-10 FIR filter channel with binary input.
The GB drawn with Gaussian distributed input provides an upper bound. The true curve can be found by the SD tree search with infinite sphere radius (i.e., $I_{\mathrm{DFS}}^{up}$ with $\alpha\rightarrow\infty$).
The SA method is the worst and the HD1 method is better than the SA method at medium SNR.
The BCJR method with full trellis and the proposed DFS-based SD upper bound with $\alpha=1.5$ provide the true curve.
The DFS-based SD upper bound with $\alpha=1$, the BFS-based SD upper bounds, and the BCJR-based RSUB with $Q=100$ yield some errors as SNR decreases.
For larger $\alpha$ and $K$ parameters, the SD upper bounds become more accurate.

Fig.~\ref{fig:CH10_all}~(b) shows the complexity in terms of the number of visited nodes during the tree search or states during the trellis search.
The HD1 method requires to find $N_t(M_c-1)$ neighbor components. The number of visited nodes in full tree search for the true curve is given by $\sum_{k=1}^{N_t}M_c^{k}$. The number of visited states in the BCJR method with full trellis at the $N_t$-th stage\footnote{For fair comparison, we consider the complexity corresponding to first $N_t$ symbols for the BCJR method since the SD bounds has the block length $N_t$.} is given by $M_c\sum_{k=0}^{N_t-1} M_c^{k}$.
Unlike the BFS-based SD upper bound and the BCJR method, the DFS-based SD upper bounds show variable complexities according to SNR due to fixed sphere radius, i.e., they result in higher complexity at low SNR. 
The complexity of the BFS-based SD upper bound is given in \eqref{eq:C_K} and that of the BCJR-based RSUB is obtained by $M_c(\sum_{k=0}^{q_0}M_c^{k} + \sum_{k=q_0+1}^{N_t-1}Q)$ where $q_0\triangleq\max\{k:M_c^{k}<Q\}$.
The BFS with $K=50$ has lower complexity than the BCJR-based RSUB with $Q=100$, while it is much more accurate on the mutual information as shown in Fig.~\ref{fig:CH10_all}~(a). Moreover, the BFS with $K=50$ is more accurate than the DFS with $\alpha=1$, while it has much lower complexity when $\mathrm{SNR}\leq4$ dB. Thus, the BFS is useful for low-complexity with a reasonable accuracy.

\begin{figure}
  \centering
  \includegraphics[width=8.5cm]{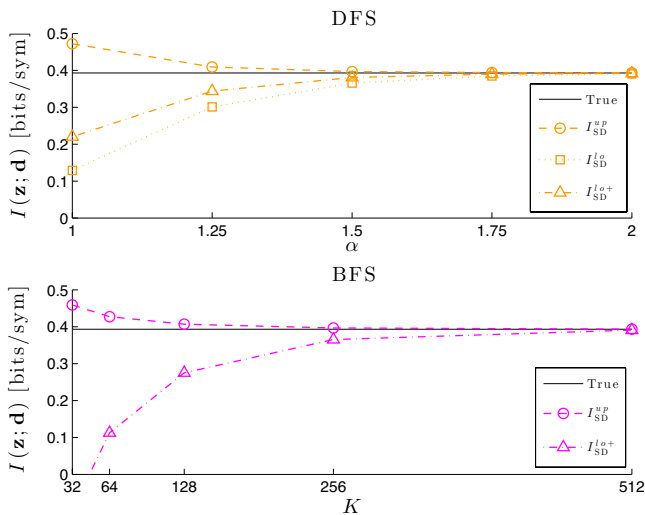}
  \caption{SD upper bound ($I_{\mathrm{x}}^{up}$), lower bound ($I_{\mathrm{x}}^{lo}$), and enhanced lower bound ($I_{\mathrm{x}}^{lo+}$) for DFS (upper figure) and BFS (lower figure) at $\mathrm{SNR}=-2.5$ dB in memory-10 FIR filter channel with binary input. }
\label{fig:CH10_up_lo}
\vspace{-0.4cm}
\end{figure}

Fig.~\ref{fig:CH10_up_lo} shows trends of the SD bounds according to control parameters (i.e., $\alpha$ for DFS and $K$ for BFS) in memory-10 FIR filter channel with binary input at $\mathrm{SNR}=-2.5$ dB. As the parameters increases, the bounds converge to the true curve.
Note that for both DFS and BFS cases, the upper bounds are much tighter than the lower and enhanced lower bounds.

\subsection{Frequency-Selective and Time-Selective Fading Channel with a Large Memory}

As second example, we consider a generalized frequency- and time-selective fading channel given by
$\mathbf{H}=\mathbf{AG}$ where $\mathbf{A}$ is the diagonal time-selective channel matrix and $\mathbf{G}$ is the frequency-selective circulant matrix as in \cite{TCS+10TIT}.  This channel setup is relevant for realistic WCDMA systems \cite{DOK+14VTC}.
For $\mathbf{A}=\mathrm{diag}(a_1, \ldots, a_{N_t})$, we assume $a_i \sim \mathcal{CN}(0,1)\forall i$.
For $\mathbf{G}$, we consider a memory-$L$ FIR filter channel, i.e., $z_k = \sum_{l=0}^{L} g_l d_{k-l} + n_k$, where $g_{l} = 2^{-l}, l=0,\ldots,L$.
Note that the BCJR method in \cite{ALV+06TIT} does not work for these setups due to time-varying property. Thus, we only take into account the SA and HD1 methods as benchmarks in this channel.

\begin{figure}
    \centering
    \subfigure[]{
        \includegraphics[width=8.5cm]{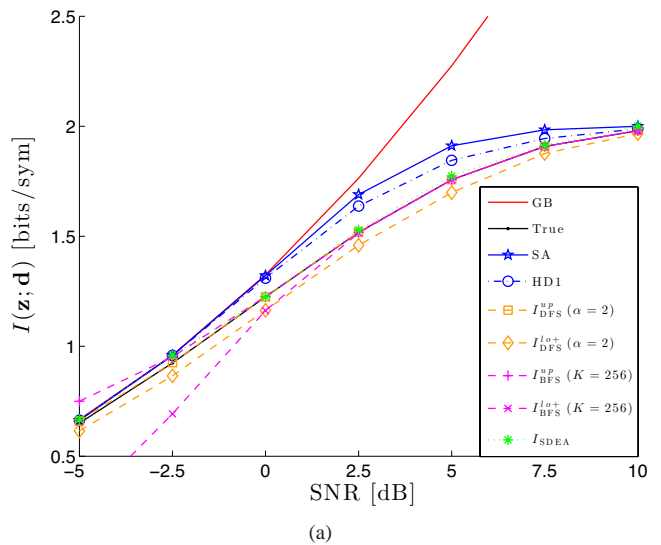}}   
    \subfigure[]{
        \includegraphics[width=8.5cm]{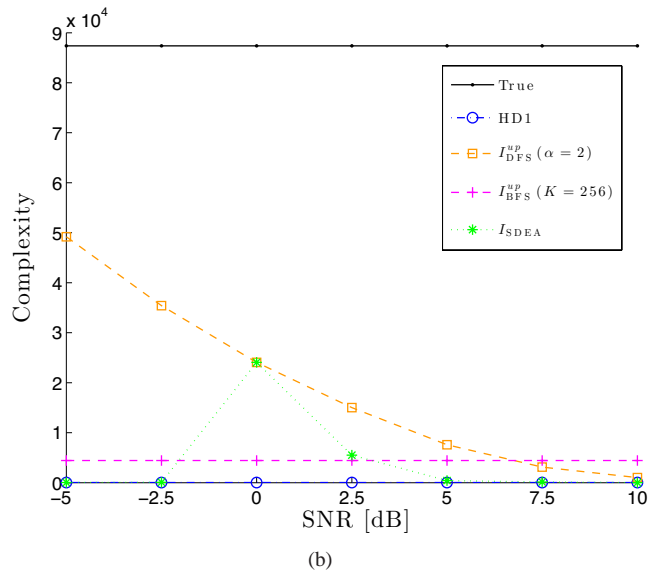}}
\caption{Frequency-selective and time-selective channel with 4-QAM input, $N_t=8$, and $L=N_t-1$ (a) Mutual information [bits/symbol] (b) Complexity in terms of average number of visited nodes during the tree search. For Monte-Carlo expectation, we use $N_d=50$ and $N_z=50$. For the enhanced approximation, we use $\alpha=2$, $10\log_{10}\gamma_l = -4$ dB, $10\log_{10}\gamma_h = 4$ dB at $\rho_c=0$ dB. Note that in the randomly realized channel $\mathbf{H}$, $10\log_{10}\lambda_1^2=-3.51$ dB and $10\log_{10}\lambda_{N_t}^2=-3.89$ dB.}
\label{fig:Nt8}
\vspace{-0.4cm}
\end{figure}

Fig.~\ref{fig:Nt8} shows the mutual information and complexity for frequency- and time-selective channel with 4-QAM input, $N_t = 8$, and $L=N_t-1$.
The DFS-based upper bound is almost the same as the true curve with much lower complexity, while the BFS-based upper bound has small errors at low SNR due to lowering the complexity. 
As investigated in Fig.~\ref{fig:CH10_up_lo}, the upper bounds are much tighter than the lower bounds.
The DFS-based lower bound has approximately a constant gap with the upper bound, whereas the BFS-based lower bound is tight at high SNR but loose at low SNR.
The enhanced approximation approaches the true curve with a small gap but much lower complexity at low and high SNRs, while the SA and HD1 methods have large errors at moderate SNR.

\begin{figure}
  \centering
  \includegraphics[width=8.5cm]{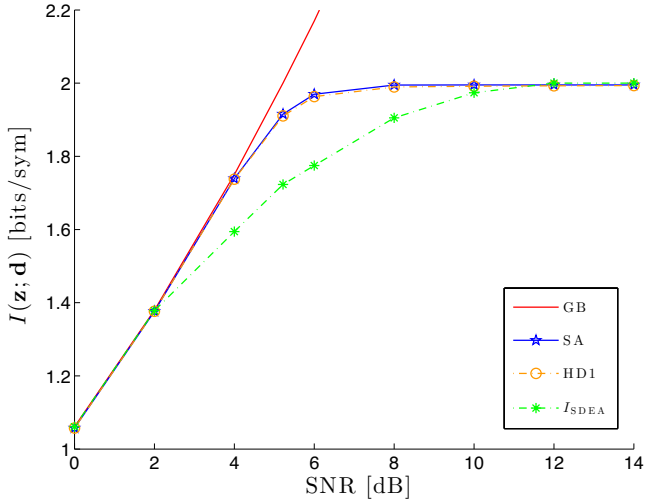}
  \caption{Mutual information in frequency-selective and time-selective channel with 4-QAM input, $N_t = 40$, and $L=N_t-1$. For Monte-Carlo expectation, we $N_d=50$ and $N_z=50$. For the enhanced approximation, we use $\alpha=1$, $10\log_{10}\gamma_l = 3$ dB, $10\log_{10}\gamma_h = 5$ dB at $\rho_c=0$ dB.}
\label{fig:Nt40}
\vspace{-0.4cm}
\end{figure}

Fig.~\ref{fig:Nt40} shows the mutual information for frequency- and time-selective channel with 4-QAM input, $N_t = 40$, and $L = N_t -1$. Computing the true curve is impossible due to the huge problem size, i.e., $M_c^{N_t}=4^{40}\approx 1.2\times 10^{24}$. The SD bounds are also unavailable within reasonable simulation time. Therefore, we compare the enhanced approximation with the SA and HD1 methods. Both the SA and HD1 methods almost reach two trivial upper bounds, i.e., GB and SEB, for this large size case, while the enhanced approximation still yields a {\it nice} curve below. 
Note that from the properties given in Section~\ref{subsec:gamma_ths} and Fig.~\ref{fig:Nt8}, we can conjecture that the true curve lies below the enhanced approximation.
The complexity of the enhanced approximation is about $10^4$ at $\mathrm{SNR}=4$ dB and less than $10^2$ in the other SNRs, while the complexity of full tree search for the true curve is $4(4^{40}-1)/3\approx 1.6\times 10^{24}$.
\REV{Compared Fig.~\ref{fig:Nt40} to Fig.~\ref{fig:Nt8}~(a), for a large block size, the mutual information is decreased in overall but in general, it will depend on the channel realization.}

\section{Conclusion} \label{sec:conclusion}

We have proposed novel complexity efficient algorithmic solutions to approximate the entropy of Gaussian mixture distributions with a large number of components. The algorithms allow to trade-off the accuracy versus the complexity and the approximations are asymptotically tight with unbounded complexity. The extended approach can even deal with very high system dimensions with a reasonable accuracy which was not possible previously. The computation of the entropy for Gaussian mixture distribution is important for many problems, e.g. data fusion, machine learning, etc. In particular, it can be used to approximate the mutual information of a vector-valued Gaussian channel with finite input alphabets. In contrast to other methods, the proposed algorithms are applicable to any linear input output relation. The proposed concepts can be easily adapted or extended to other application areas.
\REV{For future work, the concept and methods developed in this work can be extended to deal with more general Gaussian mixture distributions with heterogeneous covariance structures including improper complex signals.}


\appendices
\def\thesection{\Alph{section}}%
\def\thesectiondis{\Alph{section}}%

\section{Proofs of Properties on $\gamma_l$ and $\gamma_h$}
\setcounter{equation}{0}
\renewcommand{\theequation}{A.\arabic{equation}}
\label{app:Proof_of_Properties}

\begin{itemize}
\item[a)] Property~a corresponds to the case of $N_C=N_t$. Thus, we show that the mutual information of the Babai estimate-based approximation with $N_C=N_t$ is equivalent to $I_{\mathrm{SEB}}^{up}$. 
Instead of \eqref{eq:v_C}, the effective received signal becomes $\mathbf{v}\approx\mathbf{R}\tilde{\mathbf{d}}+\mathbf{w}$. 
Applying the single component only approximation, the pdf of $\mathbf{v}$ is given by
\begin{align}
f(\mathbf{v}) &= \frac{\exp\left(-\|\mathbf{v}-\mathbf{R}\tilde{\mathbf{d}}\|^2\right)}{(\pi M_c)^{N_t}} \approx \frac{\exp\left(-\|\mathbf{w}\|^2\right)}{(\pi M_c)^{N_t}}.\nonumber
\end{align}
Then, the mutual information is derived as
\begin{align}
I(\mathbf{z};\mathbf{d}) 
&\approx \mathbb{E}\left[\log_2(\pi M_c)^{N_t} + \log_2 e^{\|\mathbf{w}\|^2}\right] - \log_2 (\pi e)^{N_t}\nonumber\\
&= \log_2 M_c^{N_t} + \log_2 e^{\mathbb{E}\left[\|\mathbf{w}\|^2\right]} - \log_2 e^{N_t}=I_{\mathrm{SEB}}^{up}\nonumber
\end{align}
since $\mathbb{E}\left[\|\mathbf{w}\|^2\right]=N_t$ and $I_{\mathrm{SEB}}^{up}=H(\mathbf{d})=\log_2 M_c^{N_t}$.\hfill $\blacksquare$
\item[b)] Property~b corresponds to the case of $N_A=N_t$. Thus, we show that the mutual information of the single Gaussian approximation with $N_A=N_t$ is equivalent to $I_{\mathrm{GB}}^{up}$.
First of all, $I_{\mathrm{GB}}^{up}=\log_2\det(\rho\mathbf{RR}^{\mathsf{H}} +\mathbf{I})$ since $\mathbb{E}\left[\mathbf{vv}^\mathsf{H}\right]=\mathbb{E}\left[\left(\mathbf{Rd + w}\right)\left(\mathbf{Rd + w}\right)^\mathsf{H}\right]=\rho\mathbf{RR}^{\mathsf{H}} + \mathbf{I}$.
Applying the single Gaussian approximation, the pdf of $\mathbf{v}$ is
\begin{align}
f(\mathbf{v}) \approx \frac{\exp(-\mathbf{v}^{\mathsf{H}}\mathbf{K}_{\mathbf{v}}^{-1}\mathbf{v})}{\pi^{N}\det(\mathbf{K_v})},\nonumber
\end{align}
where $\mathbf{K_v}=\rho\mathbf{RR}^{\mathsf{H}} + \mathbf{I}$ since $\mathbf{A}=\mathbf{R}$. 
Hence, the mutual information is derived as
\begin{align}
I(\mathbf{z};\mathbf{d}) 
&\approx\mathbb{E}\Big[\log_2\pi^{N_t} + \log_2 \det(\mathbf{K_v}) \nonumber\\
&\quad+ \log_2 \exp\left((\mathbf{K}_{\mathbf{v}}^{-\frac{1}{2}}\mathbf{v})^{\mathsf{H}}\mathbf{K}_{\mathbf{v}}^{-\frac{1}{2}}\mathbf{v}\right)\Big] - \log_2 (\pi e)^{N_t}\nonumber\\
&=\log_2 \det(\mathbf{K_v})=I_{\mathrm{GB}}^{up},\nonumber
\end{align}
since $\mathbf{K_v}$ is Hermitian and $\mathbf{K}_{\mathbf{v}}^{-\frac{1}{2}}\mathbf{v}\sim\mathcal{CN}(0,\mathbf{I})$.\hfill $\blacksquare$
\item[c)] The proof of Property~c is straightforward since this corresponds to the case of $N_B=N_t$.\hfill $\blacksquare$
\item[d)] Since $N_A=0$ and $N_B+N_C=N_t$ for the BC curve, $f(\mathbf{v})=f(\mathbf{v}_C) f(\mathbf{v}_B|\mathbf{v}_C)$ from \eqref{eq:f_v}, where
\begin{align}
f(\mathbf{v}_C) &\approx \frac{\exp(-\|\mathbf{w}_C\|^2)}{(\pi M_c)^{N_C}},\nonumber\\
f(\mathbf{v}_B|\mathbf{v}_C) &\approx \sum_{m=1}^{|\mathcal{D}_B^{\mathrm{SD}}|}\frac{\exp(-\|\mathbf{B}(\mathbf{d}_B - \hat{\mathbf{d}}_{B,m}) + \mathbf{w}_B\|^2)}{(\pi M_c)^{N_B}},\nonumber
\end{align}
where $\hat{\mathbf{d}}_{B,m}$ is the $m$-th vector in $\mathcal{D}_B^{\mathrm{SD}}$.
Denoting $\mathbf{e}_{B,m}\triangleq \mathbf{B}(\mathbf{d}_B - \hat{\mathbf{d}}_{B,m})$, the mutual information is derived as
\begin{align}
I(\mathbf{z};\mathbf{d}) 
&\approx \mathbb{E}\Big[\log_2 (\pi M_c)^{N_C} + \log_2 e^{\|\mathbf{w}_C\|^2} + \log_2 (\pi M_c)^{N_B}\nonumber\\
&\quad - \log_2 \sum_{m=1}^{|\mathcal{D}_B^{\mathrm{SD}}|}e^{-\|\mathbf{e}_{B,m} + \mathbf{w}_B\|^2}\Big] - \log_2 (\pi e)^{N_t}\nonumber\\
&\overset{(\alpha)}{=} \log_2 M_c^{N_t} - \mathbb{E}\Big[\log_2 \sum_{m=1}^{|\mathcal{D}_B^{\mathrm{SD}}|}e^{-\|\mathbf{e}_{B,m} + \mathbf{w}_B\|^2}\Big]\nonumber\\
&\quad + \log_2 e^{\mathbb{E}[\|\mathbf{w}_C\|^2]} - \log_2 e^{\mathbb{E}[\|\mathbf{w}_B\|^2 + \|\mathbf{w}_C\|^2]}\nonumber\\
&= \underbrace{\log_2 M_c^{N_t}}_{=I_{\mathrm{SEB}}^{up}} - \mathbb{E}\Big[\log_2 \underbrace{\sum_{m=1}^{|\mathcal{D}_B^{\mathrm{SD}}|}e^{\|\mathbf{w}_B\|^2-\|\mathbf{e}_{B,m} + \mathbf{w}_B\|^2}}_{\triangleq \mathcal{X}}\Big],\label{eq:MI_d}
\end{align}
where ($\alpha$) comes from $\mathbb{E}[{\|\mathbf{w}\|^2}]= \mathbb{E}[\|\mathbf{w}_B\|^2 + \|\mathbf{w}_C\|^2]=N_t$.
If $\mathcal{X} \geq 1$, the second term of \eqref{eq:MI_d} becomes non-positive and therefore, $I(\mathbf{z};\mathbf{d}) \leq I_{\mathrm{SEB}}^{up}$ always holds.\\
If $\mathbf{d}_{B} \in \mathcal{D}_B^{\mathrm{SD}}$, $\mathcal{X}$ includes $\exp(\|\mathbf{w}_B\|^2 - \|\mathbf{e}_{B,m} + \mathbf{w}_B\|^2)=1$ and thus, $\mathcal{X}\geq 1$. If $\mathbf{d}_{B} \notin \mathcal{D}_B^{\mathrm{SD}}$, all vectors in $\mathcal{D}_B^{\mathrm{SD}}$ yields shorter Euclidean distances than $\mathbf{d}_B$. That is, for all $\hat{\mathbf{d}}_{B,m}\in\mathcal{D}_B^{\mathrm{SD}}$, $\exp(\|\mathbf{w}_B\|^2 - \|\mathbf{e}_{B,m} + \mathbf{w}_B\|^2) \geq 1$ since $\|\mathbf{w}_B\|^2 \geq \|\mathbf{e}_{B,m} + \mathbf{w}_B\|^2$. 
As a result, $\mathcal{X}\geq |\mathcal{D}_B^{\mathrm{SD}}| \geq 1$. Therefore, $I(\mathbf{z};\mathbf{d}) \leq I_{\mathrm{SEB}}^{up}$ always holds.\\
If $\gamma_h<\lambda_{1}^2$ then $I(\mathbf{z};\mathbf{d})=I_{\mathrm{SEB}}^{up}$ by Property~a. In addition, if $\gamma_h\geq\lambda_{N_t}^2$ then $I(\mathbf{z};\mathbf{d})=I_{\mathrm{SD}}^{up}$ by Property~c. As $\gamma_h$ increases from $\lambda_1^2$, $N_B$ becomes non-zero and $\mathcal{X}$ has $M_c^{N_B}$ exponential terms. For further increasing $\gamma_h$, if $N_B$ increases by one then $\mathcal{X}$ has $M_c$ additional exponential terms. Since $\exp(\cdot)\geq 0$, $\mathcal{X}$ gradually increases as $\gamma_h$ increases. Therefore, $I(\mathbf{z};\mathbf{d})$ monotonically decreases from $I_{\mathrm{SEB}}^{up}$ to $I_{\mathrm{SD}}^{up}$ as $\gamma_h$ increases.\hfill $\blacksquare$
\item[e)] Since $N_C=0$ and $N_A+N_B=N_t$ for the AB curve, $f(\mathbf{v})\approx\sum_{m=1}^{|\mathcal{D}_{B}^{\mathrm{SD}}|}p(\hat{\mathbf{d}}_{B,m})f(\mathbf{v}_B|\hat{\mathbf{d}}_{B,m}) f(\mathbf{v}_A|\hat{\mathbf{d}}_{B,m})$ in \eqref{eq:f_v}.
Denoting $\mathbf{e}_{B,m}\triangleq \mathbf{B}(\mathbf{d}_B - \hat{\mathbf{d}}_{B,m})$ and $\mathbf{v}_{A,m} \triangleq \mathbf{v}_A-\mathbf{B}_A \hat{\mathbf{d}}_{B,m}=\mathbf{A}\mathbf{d}_A + \mathbf{B}_{A}(\mathbf{d}_B - \hat{\mathbf{d}}_{B,m}) + \mathbf{w}_A$, the pdfs are written by $p(\hat{\mathbf{d}}_{B,m}) = \frac{1}{M_c^{N_B}}$,
\begin{align}
f(\mathbf{v}_B|\hat{\mathbf{d}}_{B,m}) &= \frac{\exp(-\|\mathbf{e}_{B,m}+\mathbf{w}_B\|^2)}{\pi^{N_B}},\nonumber\\
f(\mathbf{v}_A|\hat{\mathbf{d}}_{B,m}) &= \frac{\exp(-\mathbf{v}_{A,m}^{\mathsf{H}}\mathbf{K}_{A}^{-1}\mathbf{v}_{A,m})}{\pi^{N_A}\det(\mathbf{K}_{A})},\nonumber
\end{align}
where $\mathbf{K}_A=\rho\mathbf{AA}^{\mathsf{H}} + \mathbf{I}$.
Then, the mutual information is derived as
\begin{align}
&I(\mathbf{z};\mathbf{d}) 
\approx -\mathbb{E}\Big[\sum_{m=1}^{|\mathcal{D}_{B}^{\mathrm{SD}}|}p(\hat{\mathbf{d}}_{B,m})f(\mathbf{v}_B|\hat{\mathbf{d}}_{B,m}) f(\mathbf{v}_A|\hat{\mathbf{d}}_{B,m})\Big]\nonumber\\
&\quad\quad\quad\quad - \log_2 (\pi e)^{N_t}\nonumber\\
&\overset{(\beta)}{=} \mathbb{E}\Big[ \log_2 M_c^{N_B} + \log_2 \pi^{N_A+N_B} + \log_2\det(\mathbf{K}_A) \nonumber\\
&\quad\quad - \log_2 \sum_{m=1}^{|\mathcal{D}_B^{\mathrm{SD}}|}e^{-\|\mathbf{e}_{B,m}+\mathbf{w}_B\|^2 -\mathbf{v}_{A,m}^{\mathsf{H}}\mathbf{K}_{A}^{-1}\mathbf{v}_{A,m}}\Big] \nonumber\\
&\quad\quad - \log_2 \pi^{N_t} - \mathbb{E}\left[\log_2 e^{\|\mathbf{w}\|^2}\right]\nonumber
\end{align}
\begin{align}
&\overset{(\delta)}{=} \log_2 M_c^{N_B}  + \log_2 \det(\rho\mathbf{A}\mathbf{A}^{\mathsf{H}}+\mathbf{I}) - \mathbb{E}\Big[\log_2 \nonumber\\
&\quad\quad \underbrace{\sum_{m=1}^{|\mathcal{D}_B^{\mathrm{SD}}|}e^{\|\mathbf{w}_B\|^2-\|\mathbf{e}_{B,m}+\mathbf{w}_B\|^2+\|\mathbf{w}_A\|^2 - \mathbf{v}_{A,m}^{\mathsf{H}}\mathbf{K}_{A}^{-1}\mathbf{v}_{A,m}}}_{\triangleq\mathcal{Y}}\Big]\nonumber
\end{align}
where ($\beta$) comes from $\mathbb{E}[{\|\mathbf{w}\|^2}]=N_t$, and ($\delta$) comes from $\|\mathbf{w}\|^2 = \|\mathbf{w}_A\|^2+\|\mathbf{w}_B\|^2$.
As $\rho\rightarrow 0$, $\log_2 \det(\rho\mathbf{A}\mathbf{A}^{\mathsf{H}}+\mathbf{I})\approx 0$ and we have
\begin{align}
\mathcal{Y} &\approx \sum_{m=1}^{|\mathcal{D}_B^{\mathrm{SD}}|}e^{\|\mathbf{w}_B\|^2-\|\mathbf{e}_{B,m}+\mathbf{w}_B\|^2 + \|\mathbf{w}_A\|^2 - \|\mathbf{w}_A\|^2}= \mathcal{X} \overset{(\phi)}{\geq} 1,\nonumber
\end{align}
where $(\phi)$ comes from the proof of Property d. Therefore, for $\rho\rightarrow 0$, we have $I(\mathbf{z};\mathbf{d}) = \log_2 M_c^{N_B} - \mathbb{E}[\log_2\mathcal{X}]\geq \log_2 M_c^{N_B} - \log_2\sum_{m=1}^{|\mathcal{D}_B^{\mathrm{SD}}|}e^{\mathbb{E}[\|\mathbf{w}_B\|^2]}=\log_2 M_c^{N_B} - (\log_2 |\mathcal{D}_B^{\mathrm{SD}}|+{N_B\log_2 e})\geq -{N_B\log_2 e}$, which can be positive, while $I_{\mathrm{GB}}^{up} = \log_2\det(\rho\mathbf{R}\mathbf{R}^{\mathsf{H}} + \mathbf{I})\approx 0$ as $\rho\rightarrow 0$. 
Therefore, the AB curve can exceed $I_{\mathrm{GB}}^{up}$ at low SNR.
\hfill $\blacksquare$
\end{itemize}

\balance

\bibliographystyle{./style/IEEEtran_v111}
\bibliography{./style/IEEEabrv,./style/RefAbrv,Reference}

\begin{thebibliography}{10}
\providecommand{\url}[1]{#1}
\csname url@rmstyle\endcsname
\providecommand{\newblock}{\relax}
\providecommand{\bibinfo}[2]{#2}
\providecommand\BIBentrySTDinterwordspacing{\spaceskip=0pt\relax}
\providecommand\BIBentryALTinterwordstretchfactor{4}
\providecommand\BIBentryALTinterwordspacing{\spaceskip=\fontdimen2\font plus
\BIBentryALTinterwordstretchfactor\fontdimen3\font minus
  \fontdimen4\font\relax}
\providecommand\BIBforeignlanguage[2]{{%
\expandafter\ifx\csname l@#1\endcsname\relax
\typeout{** WARNING: IEEEtran.bst: No hyphenation pattern has been}%
\typeout{** loaded for the language `#1'. Using the pattern for}%
\typeout{** the default language instead.}%
\else
\language=\csname l@#1\endcsname
\fi
#2}}

\bibitem{HBD+08MFI}
{M. F. Huber, T. Bailey, H. Durrant-Whyte, and U. D. Hanebeck}, ``On entropy
  approximation for {G}aussian mixture random vectors,'' in \emph{Proc. IEEE
  Int'l Conf. Multisensor Fusion and Integration for Intelligent Systems
  (MFI)}, Aug. 2008.

\bibitem{GVR14TWC}
{M. A. Girnyk, M. Vehkaper\"{a}, and L. Rasmussen}, ``Large-system analysis of
  correlated {MIMO} channels with arbitrary signaling in the presence of
  interference,'' \emph{{IEEE} Trans. Wireless Commun.}, vol.~13, no.~4, pp.
  2060 -- 2073, Apr 2014.

\bibitem{ZSF+03WOC}
{H. Zhu, Z. Shi, B. Farhang-Beroujeny, and C. Schlegel}, ``An efficient
  statistical approach for calculation of capacity of {MIMO} channels,'' in
  \emph{Proc. Wirel. Opt. Commun. (WOC)}, July 2003, Online available:
  www.ece.ualberta.ca/~hcdc/Library/ZhuShiFarSch03.pdf.

\bibitem{ALV+06TIT}
{D. M. Arnold, H.-A. Loeliger, P. O. Vontobel, A. Kav\v{c}i\'{c}, and W. Zeng},
  ``Simulation-based computation of information rates for channels with
  memory,'' \emph{{IEEE} Trans. Inf. Theory}, vol.~52, no.~8, pp. 3498--3508,
  Aug. 2006.

\bibitem{DOK+14VTC}
{T. T. Do, T. J. Oechtering, S. M. Kim, and G. Peters}, ``Capacity analysis of
  continuous-time time-variant asynchronous uplink {WCDMA} system,'' in
  \emph{Proc. IEEE Veh. Tech. Conf. (VTC)}, Sept. 2014.

\bibitem{CWP+11Fusion}
{D. F. Crouse, P. Willett, K. Pattipati, and L. Svensson}, ``A look at
  {G}aussian mixture reduction algorithms,'' in \emph{Proc. IEEE Int'l Conf.
  Inf. Fusion (FUSION)}, July 2011.

\bibitem{HuH08Fusion}
{M. F. Huber and U. D. Hanebeck}, ``Progressive {G}aussian mixture reduction,''
  in \emph{Proc. IEEE Int'l Conf. Inf. Fusion (FUSION)}, June 2008.

\bibitem{ScH09Fusion}
{D. Schieferdecker and M. F. Huber}, ``{G}aussian mixture reduction via
  clustering,'' in \emph{Proc. IEEE Int'l Conf. Inf. Fusion (FUSION)}, July
  2009.

\bibitem{CGJ96JAIR}
{D. A. Cohn, Z. Ghahramani, and M. I. Jordan}, ``Active learning with
  statistical models,'' \emph{Journal of Artificial Intelligence Research},
  vol.~4, pp. 129--145, Mar. 1996.

\bibitem{NG08TM}
{A. Nikseresht and M. Gelgon}, ``Cossip-based computatin of a {G}aussian
  mixture model for distributed multimedia indexing,'' \emph{{IEEE} Trans.
  Multimedia}, vol.~10, no.~3, pp. 385--392, Apr. 2008.

\bibitem{AK12LNCS}
{Y. Avrithis and Y. Kalantidis}, ``Approximate gaussian mixtures for large
  scale vocabularies,'' \emph{Lecture Notes in Computer Science}, vol. 7574,
  pp. 15--28, 2012.

\bibitem{BGP10PR}
{P. Bruneau, M. Gelgon, and F. Picarougne}, ``Parsimonious reduction of
  {G}aussian mixture models with a variational-{B}ayes approach,''
  \emph{Pattern Recognition}, vol.~43, no.~3, pp. 850--858, Mar. 2010.

\bibitem{Sal09TAES}
{D. J. Salmond}, ``Mixture reduction algorithms for point and extended object
  tracking in clutter,'' \emph{{IEEE} Trans. Aerosp. Electron. Syst.}, vol.~45,
  no.~2, pp. 667--686, Apr. 2009.

\bibitem{Run07TAES}
{A. R. Runnalls}, ``{Kullback-Leibler} approach to {Gaussian} mixture
  reduction,'' \emph{{IEEE} Trans. Aerosp. Electron. Syst.}, vol.~43, no.~3,
  pp. 989--999, July 2007.

\bibitem{DL08TIT}
{J. Dauwels and H.-A. Loeliger}, ``Computation of information rates by particle
  methods,'' \emph{{IEEE} Trans. Inf. Theory}, vol.~54, no.~1, pp. 406--409,
  Jan. 2008.

\bibitem{ML13TIT}
{M. Molkaraie and H.-A. Loeliger}, ``Monte carlo algorithms for the partition
  function and information rates of two-dimensional channels,'' \emph{{IEEE}
  Trans. Inf. Theory}, vol.~59, no.~1, pp. 495--503, Jan. 2013.

\bibitem{FiP85MC}
{U. Fincke and M. Pohst}, ``Improved methods for calculating vectors of short
  length in lattice, including a complexity analysis,'' \emph{Math. Comput.},
  vol.~44, pp. 463--471, Apr. 1985.

\bibitem{AEV+02TIT}
{E. Agrell, T. Eriksson, A. Vardy, and K. Zeger}, ``Closest point search in
  lattices,'' \emph{{IEEE} Trans. Inf. Theory}, vol.~48, no.~8, pp. 2201--2214,
  Aug. 2002.

\bibitem{DGC03TIT}
{M. O. Damen, H. El Gamal, and G. Caire}, ``On maximum-likelihood detection and
  the search for the closest lattice point,'' \emph{{IEEE} Trans. Inf. Theory},
  vol.~49, no.~10, pp. 2389--2402, Oct. 2003.

\bibitem{MGD+06TIT}
{A. D. Murugan, H. El Gamal, M. O. Damen, and G. Caire}, ``A unified framework
  for tree search decoding: rediscovering the sequential decoder,''
  \emph{{IEEE} Trans. Inf. Theory}, vol.~52, no.~3, pp. 933--953, Mar. 2006.

\bibitem{HaV05TSP}
{B. Hassibi and H. Vikalo}, ``On the sphere-decoding algorithm {I}. expected
  complexity,'' \emph{{IEEE} Trans. Signal Process.}, vol.~53, no.~8, pp.
  2806--2818, Aug. 2005.

\bibitem{ViH05TSP}
{H. Vikalo and B. Hassibi}, ``On the sphere-decoding algorithm {II}.
  generalizations, second-order statistics, and applications to
  communications,'' \emph{{IEEE} Trans. Signal Process.}, vol.~53, no.~8, pp.
  2806--2818, Aug. 2005.

\bibitem{JaO05TSP}
{J. Jald\'{e}n and B. Ottersten}, ``On the complexity of sphere decoding in
  digital communications,'' \emph{{IEEE} Trans. Signal Process.}, vol.~53,
  no.~4, pp. 1474--1484, Apr. 2005.

\bibitem{GuN06JSAC}
{Z. Guo and P. Nilsson}, ``Algorithm and implementation of the $k$-best sphere
  decoding for {MIMO} detection,'' \emph{{IEEE} J. Sel. Areas Commun.},
  vol.~24, no.~3, pp. 491--503, Mar. 2006.

\bibitem{BaT08TWC}
{L. G. Barbero and J. S. Thompson}, ``Fixing the complexity of the sphere
  decoding for {MIMO} detection,'' \emph{{IEEE} Trans. Wireless Commun.},
  vol.~7, no.~6, pp. 2131--2142, June 2008.

\bibitem{BGB+03GC}
{J. Boutros, N. Gresset, L. Brunel, and M. Fossorier}, ``Soft-input soft-output
  lattice sphere decoder for linear channels,'' in \emph{Proc. {IEEE}
  {GLOBECOM}}, Dec. 2003.

\bibitem{Gal68Wil}
{R. G. Gallager}, \emph{Information Theory and Reliable Communication}.\hskip
  1em plus 0.5em minus 0.4em\relax Wiley, 1968.

\bibitem{WTC+02ISCS}
{K.-W. Wong, C.-Y. Tsui, R. S.-K. Cheng, and W.-H. Mow}, ``A {VLSI}
  architecture of a {K}-best lattice decoding algorithm for {MIMO} channels,''
  in \emph{Proc. IEEE Int'l Symp. Circuits Syst.}, May 2002.

\bibitem{RBO04ICC}
{D. L. Ruyet, T. Bertozzi, and B. \"{O}zbek}, ``Breadth first algorithms for
  {APP} detectors over {MIMO} channels,'' in \emph{Proc. {IEEE} {ICC}}, June
  2004.

\bibitem{Babai}
{M. Grotschel, L. Lov\'{a}sz, and A. Schriver}, \emph{Geometric Algorithms and
  Combinatorial Optimization, 2nd ed.}\hskip 1em plus 0.5em minus 0.4em\relax
  New York: Springer-Verlag, 1993.

\bibitem{KFL01TIT}
{F. R. Kschischang, B. J. Frey, and H.-A. Loeliger}, ``Factor graphs and the
  sum-product algorithm,'' \emph{{IEEE} Trans. Inf. Theory}, vol.~47, no.~2,
  pp. 498--519, Feb. 2001.

\bibitem{TCS+10TIT}
{A. M. Tulino, G. Caire, S. Shamai, and S. Verd\'{u}}, ``Capacity of channels
  with frequency-selective and time-selective fading,'' \emph{{IEEE} Trans.
  Inf. Theory}, vol.~56, no.~3, pp. 1187--1215, Mar. 2010.

\end{thebibliography}

\end{document}